\documentclass[twocolumn,pre,aps,superscriptaddress,nofootinbib]{revtex4-2}
\usepackage[utf8]{inputenc}

\usepackage{graphicx}
\usepackage{dcolumn}
\usepackage{gensymb}
\usepackage{svg}
\usepackage{xcolor}
\usepackage{amsmath,amssymb,dsfont,amstext,amsfonts}
\usepackage[colorlinks=true]{hyperref}  
\usepackage{bigints}

\usepackage[capitalize]{cleveref}   
\usepackage{bm}
\usepackage{physics}

\usepackage{appendix}
\hypersetup{
    bookmarks=true,         
    unicode=false,          
    pdftoolbar=true,        
    pdfmenubar=true,        
    pdffitwindow=false,     
    pdfstartview={FitH},    
    pdfsubject={},   
    pdfcreator={},   
    pdfproducer={}, 
    pdfkeywords={} {} {}, 
    pdfnewwindow=true,      
    colorlinks=true,       
    linkcolor=blue, 
    citecolor=blue,        
    filecolor=magenta,      
    urlcolor=blue,           
} 

\begin{document}
\title{On the origin of energy gaps in quasicrystalline potentials}
\author{Emmanuel Gottlob}
\email{emg69@cantab.ac.uk}
 \affiliation{Cavendish Laboratory, University of Cambridge, JJ Thomson Avenue, Cambridge CB3 0US, United Kingdom}
\author{David Gröters}
\email{david.groeters@mpq.mpg.de}
\affiliation{Cavendish Laboratory, University of Cambridge, JJ Thomson Avenue, Cambridge CB3 0US, United Kingdom}
\affiliation{Max-Planck-Institut f\"ur Quantenoptik, 85748 Garching, Germany}
\affiliation{Munich Center for Quantum Science and Technology (MCQST), 80799 Munich, Germany}
\affiliation{Fakult\"at f\"ur Physik, Ludwig-Maximilians-Universit\"at M\"unchen, 80799 M\"unchen, Germany}

\author{Ulrich Schneider}
\email{uws20@cam.ac.uk}
\affiliation{Cavendish Laboratory, University of Cambridge, JJ Thomson Avenue, Cambridge CB3 0US, United Kingdom}

\date{\today}

\begin{abstract}
Quasicrystals, structures that are ordered yet aperiodic, defy conventional band theory, confining most studies to finite-size real-space numerics. We overcome this limitation with a configuration-space framework that predicts and explains the positions and origins of energy gaps in quasicrystalline potentials. We find that a hierarchy of gaps stems from resonant hybridization between increasingly distant neighboring sites, pinning the integrated density of states below these gaps to specific irrational areas in configuration space. Large-scale simulations of a lowest-band tight-binding model built from localized Wannier functions show excellent agreement with these predictions. By moving beyond finite-size numerics, this study advances the understanding of quasicrystalline potentials, paving the way for new explorations of their quantum properties in the infinite-size limit.

\end{abstract}
\maketitle

Quasicrystals are a fascinating class of materials at the border between periodic and disordered matter. Their self-similar structure exhibits perfect long-range correlations yet no periodicity. First discovered by Schechtmann in 1984 \citep{shechtmanMetallicPhaseLongRange1984}, who observed the ten-fold symmetric diffraction pattern of an Al-Mn alloy, quasicrystals violate the crystallographic restriction theorem which limits periodic structures to two-, three-, four-, or six-fold rotational symmetries. Since their initial discovery, quasicrystals have been synthesized by fast cooling of various alloys \citep{thielQuasicrystalsReachingMaturity1999a}, and have also been found in the remains of the Trinity nuclear explosion \citep{bindiAccidentalSynthesisPreviously2021}, meteorites \citep{bindiEvidenceExtraterrestrialOrigin2012a, agrosiNaturallyOccurringAlCuFeSi2024}, and next to a downed power line on a sand dune \citep{bindiElectricalDischargeTriggers2023}. Quasicrystals can also be engineered in synthetic settings, such as as multilayered graphene systems \citep{yaoEdgeStatesTopological2018, ahnDiracElectronsDodecagonal2018, uriSuperconductivityStrongInteractions2023}. or photonic systems \citep{Steurer2007,Vardeny2013, kraus2012topological, krausFourDimensionalQuantumHall2013, Ozawa2019, verbinTopologicalPumpingPhotonic2015a, Goblot2020}.

Quasicrystalline (QC) potentials, despite lacking true periodicity, exhibit a weaker form of repetition where any local pattern appears in approximate form infinitely many times throughout an infinite system. This property is closely connected to the mathematics of discrete aperiodic tilings, where local structural motifs repeat perfectly but without giving rise to global periodicity. The tension between locally (quasi-) disordered structure and globally almost repeating patterns endows quasicrystals with unique quantum properties. QC potentials can often be obtained as a "cut" made through a higher-dimensional periodic parent potential \citep{Senechal1995}. As a consequence, they can inherit topological properties usually seen in higher-dimensional systems, a famous example being the one-dimensional Aubry André model inheriting Chern numbers \citep{bellissardGapLabellingTheorems1992}, which which for periodic systems only exist in even dimensions. In addition, their single-particle eigenstates can undergo localisation transitions \citep{sbrosciaObservingLocalization2D2020, DevakulAnderson2017} even in one dimension \citep{aubry1980annals}, possess fractal eigenstates \citep{OstlundSchrodinger1983, YouFibonacci1991, HanCritical1994} and produce mixed spectra where extended and localized states are interweaved \citep{szaboMixedSpectraPartially2020}.

Interest in quasicrystal physics was recently renewed due to the advent of synthetic quasicrystalline quantum systems. By stacking two layers of graphene with a 30 degree relative twist angle, researchers were able to create a dodecagonal quasicrystalline lattice that hosts Dirac fermions \citep{yaoEdgeStatesTopological2018, ahnDiracElectronsDodecagonal2018}. In another recent experiment, a quasicrystal was formed by the superposition of three twisted layers of graphene \citep{uriSuperconductivityStrongInteractions2023}. Given the complexity associated with condensed-matter systems, such as defects, lattice vibrations or  sample purity, recent experiments turned to optical lattices to engineer synthetic, defect-free, quasicrystalline potentials. By shining intense laser light at specific angles onto ultracold atomic clouds, 
 these \textit{optical quasicrystals} serve as highly tunable \textit{quantum simulators} for quasicrystalline physics  \citep{maceQuantumSimulation2D2016,viebahnMatterWaveDiffractionQuasicrystalline2019, sbrosciaObservingLocalization2D2020, koschMultifrequencyOpticalLattice2022, Yu2024}.

Due to their localized single-particle eigenstates at higher potential depths, quasiperiodic systems can host many-body phases typically found in disordered systems, for instance the Bose glass phase, which can be understood as an incoherent collection of  compressible local superfluid puddles.  Recent numerical studies predicted the existence of the Bose glass phase for cold atoms in an optical quasicrystal at weak \citep{johnstoneMeanfieldPhasesUltracold2019, Johnstone2021} and strong interactions \citep{gautierStronglyInteractingBosons2021, ciardiFinitetemperaturePhasesTrapped2022, zhuThermodynamicPhaseDiagram2023}. In a recent experiment using ultracold bosons, the superfluid to Bose glass phase transition was observed in the weak interaction regime of an eight-fold optical quasicrystal \citep{Yu2024}.

Bloch's theorem does not apply to quasicrystalline potentials, making periodic band structure theory inapplicable to investigate their properties. As a consequence of this aperiodicity, most studies rely on finite size real-space numerics, or approximate the system with periodic approximants equipped with periodic boundary conditions \citep{Senechal1995, szaboNonpowerlawUniversalityOnedimensional2018}.
It is therefore challenging to make definitive conclusions about the properties of quasicrystalline potentials in the infinite-size limit. In particular, a rigorous theory explaining the origin of  energy gaps in quasicrystalline potentials, whose counterpart for periodic potentials is a cornerstone of condensed matter theory, is lacking. 

In this work, we show the existence and explain the origin of many true energy gaps, and predict the associated integrated density of states (IDoS) in the single-particle energy spectrum of quasicrystalline potentials. To describe the quasicrystalline potential in the infinite-size limit, we use our recently developed language of configuration space for quasicrystalline potentials \citep{gottlobHubbardModelsQuasicrystalline2023}, which sorts lattice sites in terms of their local environments. We show that energy gaps arise due to resonant hybridisation between neighboring sites. Focusing on the eight-fold rotationally
symmetric quasicrystalline potential (8QC) \citep{viebahnMatterWaveDiffractionQuasicrystalline2019}, our model predicts that the integrated density of states at the main gaps in the system corresponds to irrational areas in configuration space that are governed by the silver ratio $\frac{1}{1+\sqrt{2}}$. To verify our predictions, we use our recently developed techniques \citep{gottlobHubbardModelsQuasicrystalline2023} to construct  a large-scale numerical tight-binding Hamiltonian of the lowest band of the system and find excellent agreement between the numerics and our analytical predictions. 

In addition, we establish the 8QC as an ideal candidate for hosting  two-dimensional (2D) many-body localised (MBL) phases. 
In randomly disordered systems in higher dimensions, MBL is argued to be unstable due to the existence of rare Griffiths regions with small local disorder that could destabilize the MBL phase by acting as nucleating regions of thermalization avalanches \citep{deroeckStabilityInstabilityDelocalization2017}. 
Thanks to their lack of rare, low-disorder regions, quasiperiodic potentials such as the 8QC have long been considered as potential candidates for hosting the many-body localized (MBL) phase in two dimensions~\citep{agrawalQuasiperiodicManybodyLocalization2022, Crowley2022}. 
This discussion received renewed attention with the discovery of \textit{weakly-modulated lines}, that is lines of sites with arbitrarily low disorder, in the two-dimensional Aubry-André model \citep{szaboMixedSpectraPartially2020,johnstoneMeanfieldPhasesUltracold2019, strkaljCoexistenceLocalizationTransport2022, Johnstone2022, duncanCritical2024}. 
Here we show that the 8QC does not contain weakly modulated lines, and is hence an ideal  candidate for a  potential two-dimensional MBL phase.

Although the present manuscript focuses on the eight-fold rotationally symmetric quasicrystalline potential (8QC) \citep{viebahnMatterWaveDiffractionQuasicrystalline2019}, the method can be directly extended to any other quasiperiodic system with a configuration-space description.
Our results and methodology open the way to further avenues of research. The existence of true energy gaps (as opposed to pseudo-gaps, where the density of states is strongly suppressed but not exactly zero) with associated irrational integrated densities of states raises the possibility of insulating phases with irrational fillings in the system, such as fermionic band insulators or Mott insulators. Our work establishes the configuration-space representation as a useful tool for deriving analytical results about infinite-size quasicrystalline potentials and thereby brings our understanding of quasicrystalline systems closer to that of their periodic counterpart.   

The manuscript is organized as follows: in \cref{sec:8QC}, we introduce the eight-fold quasicrystalline potential, summarize its known properties, and numerically construct a large-scale tight-binding Hamiltonian of its lowest band by using our recent method for computing localized Wannier functions in non-periodic potentials. In \cref{sec:spectral_localisation}, we numerically compute the single-particle energy spectrum, and the localisation properties of the eigenstates. In \cref{sec:confighubbard}, we address the system's infinite-size limit by re-expressing the real-space tight-binding model in configuration space, where lattice sites are sorted according to the configuration of the potential in their local environment. Afterwards, in \cref{sec:hybridisation}, we show how configuration space helps to understand the localisation properties of the eigenstates in the infinite-size limit in terms of resonant lines and demonstrate that the 8QC does not contain weakly modulated lines. Finally, in \cref{sec:8QCspectrumconfig}, we predict that the intersection of the resonant lines in configuration space can give rise to global energy gaps due to the hybridisation of resonant sites. Importantly, our model predicts that the (irrational) areas enclosed by the resonant lines directly correspond to the integrated density of states below the corresponding energy gaps, and we obtain excellent agreement with numerical results. Finally, we discuss the consequences of our results and outline future avenues of inquiry in \cref{sec:conclusion}.

\section{Eight-fold quasicrystalline potential: tight-binding Hamiltonian} \label{sec:8QC}

\begin{figure}
  \centering
  \includegraphics[]{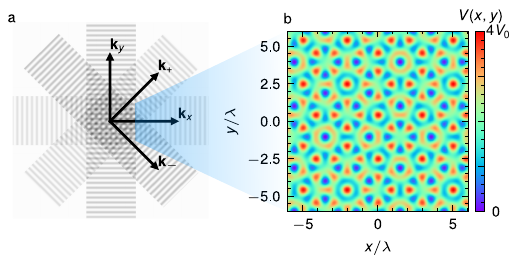}
  \caption{\textbf{Eight-fold quasicrystalline potential:} (a) The eight-fold potential is formed by superimposing two square periodic lattices ($\mathbf{k}_x, \mathbf{k}_y$ and  $\mathbf{k}_+, \mathbf{k}_-$) with a $45^\circ$ angle between them, see \cref{eq:QCpotential}. (b) Resulting potential.}
  \label{fig:QClattice}
\end{figure}

 The 8QC is formed by superimposing  four independent and mutually incoherent standing waves, see \cref{fig:QClattice}, and the resulting potential is proportional to the sum of the four intensities:
\begin{align}  \label{eq:QCpotential}
     V(\mathbf{r}) &= V_0 \sum_{i=x,y,+,-} \sin^2(\mathbf{k}_i\cdot \mathbf{r}+\phi_i) \nonumber \\
    & \mathbf{k}_i \in \frac{2\pi}{\lambda}\left\{ \begin{pmatrix} 1 \\ 0 \end{pmatrix} \,, \begin{pmatrix} 0 \\ 1 \end{pmatrix} \,, \frac{1}{\sqrt{2}} \begin{pmatrix} 1 \\ 1 \end{pmatrix} \,, \frac{1}{\sqrt{2}} \begin{pmatrix} 1 \\ -1 \end{pmatrix} \right\} 
\end{align}
where $V_0$ is the lattice depth and $\mathbf{k}_i$ and $\phi_i$ are the wave vectors and relative phases of the four individual lattices formed by counterpropagating beams of wavelength $\lambda$.

This potential is closely related to the discrete eight-fold Ammann-Beenker lattice \citep{jagannathanEightfoldOpticalQuasicrystal2013, maceQuantumSimulation2D2016}, a discrete quasicrystalline lattice constisting of squares and rhombuses.
While the potential in \cref{eq:QCpotential} is entirely deterministic and thereby clearly long-range ordered, it cannot be periodic, since its eight-fold rotational symmetry violates the crystallographic restriction theorem which states that the only allowed rotational symmetries for periodic structures are two-, three-, four-, and six-fold. Therefore, it must be quasicrystalline.

The relative phases $\phi_i$ form a $4$-dimensional parameter space, in which two orthogonal directions account for $x,y$ translations and two for phasonic degrees of freedom (which cause long-range, correlated, re-arrangements of  lattice sites) \citep{jagannathan2024propertiesammannbeenkertilingsquare}. Translating the origin by $\Delta x, \Delta y$ corresponds to the following phase shifts $\Delta\phi_i$:
\begin{equation} \label{eq:8QCtranslation}
    \begin{pmatrix}
        \Delta\phi_1 \\
        \Delta\phi_2 \\
        \Delta\phi_3 \\
        \Delta\phi_4
    \end{pmatrix} = -\frac{2\pi}{\lambda}\begin{pmatrix}
        1 & 0 \\
         0 & 1 \\
        \frac{1}{\sqrt{2}} & \frac{1}{\sqrt{2}} \\
                \frac{1}{\sqrt{2}} & - \frac{1}{\sqrt{2}} 

    \end{pmatrix}
    \begin{pmatrix}
        \Delta x \\
        \Delta y
    \end{pmatrix}
\end{equation}

The two remaining directions in the four-dimensional parameter space spanned by the phases $\phi_i$, associated with the phasonic degrees of freedom, are central in the context of topological adiabatic pumping \citep{gottlob2024quasiperiodicityprotectsquantizedtransport, marraTopologicallyQuantizedCurrent2020}. For sufficiently large system size, the precise values of the phases $\phi_i$ do not impact the average geometry of the lattice sites and hence don't change the static bulk properties of the 8QC, as was shown in \citep{koschMultifrequencyOpticalLattice2022} and \citep{gottlobHubbardModelsQuasicrystalline2023}.

The 8QC exhibits quantum properties typical of quasicrystalline systems. Its single-particle energy spectrum undergoes an energy-dependent localisation transition under increasing the lattice depth $V_0$ \citep{gottlobEmmgottlobQuasiHubbard2024a, zhuConstructionMaximallyLocalized2017}, which was verified experimentally for its ground state \citep{sbrosciaObservingLocalization2D2020}. In the presence of interactions, its phase diagram exhibits a superfluid, a Bose glass, and a Mott insulating phase \citep{johnstoneMeanfieldPhasesUltracold2019, Johnstone2021,gautierStronglyInteractingBosons2021, zhuThermodynamicPhaseDiagram2023}. Recently, the Bose glass phase was detected experimentally using ultracold atoms and it was shown that it could not be traversed adiabatically~\citep{Yu2024}, a property related to being non-ergodic. 

In what follows, we rescale all lengths and energies in the natural units describing ultracold atoms in the potential: the lattice beams wavelength $\lambda$ and the recoil energy $E_r = \frac{\hbar^2 k_0^2}{2m}$, with $k_0 = 2\pi / \lambda$ and $m$ the mass of the considered atomic species.

\begin{figure}
    \centering
    \includegraphics[width = \linewidth]{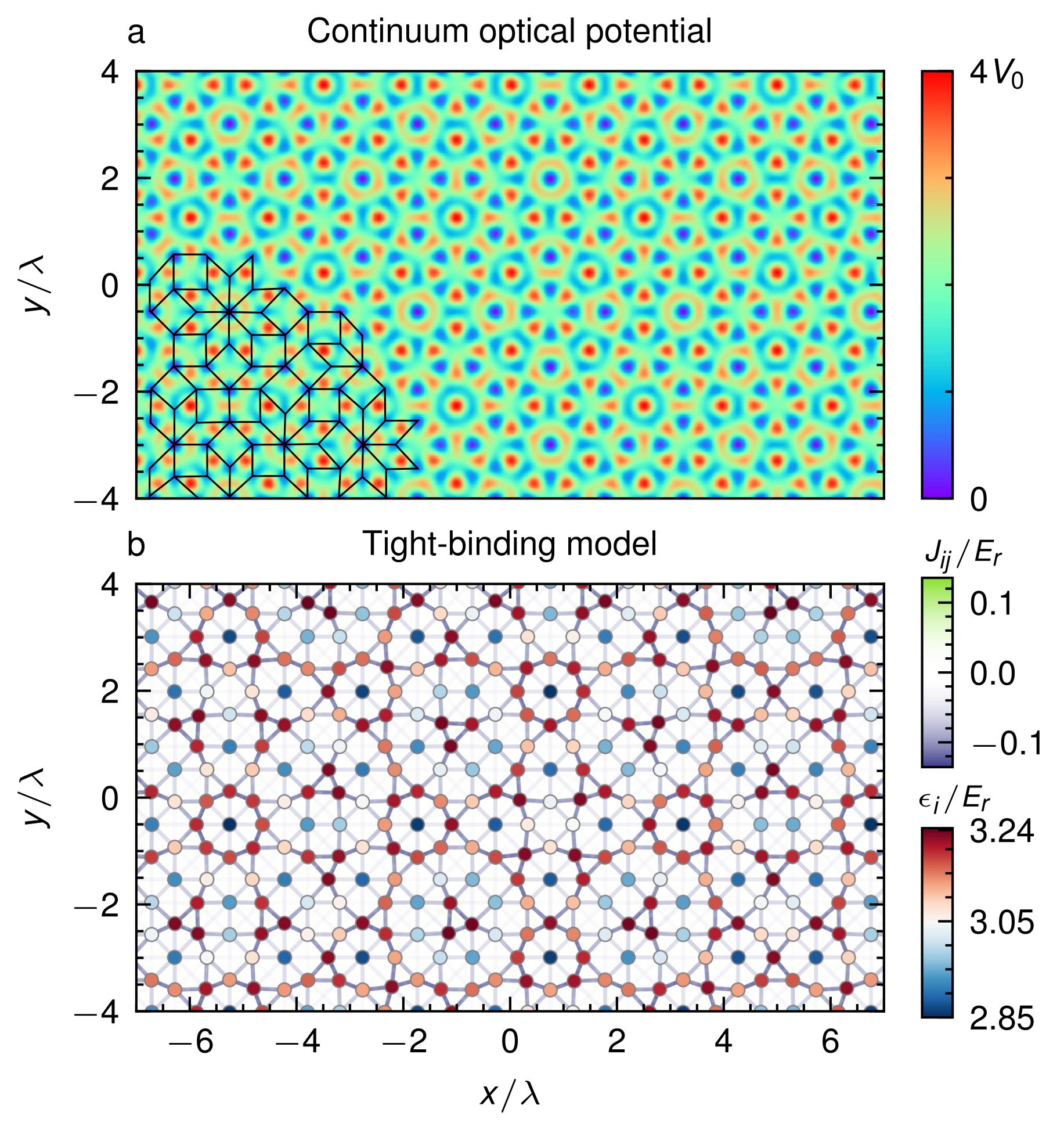}
    \caption{\textbf{Tight-binding Hamiltonian of the 8QC lowest band:} 
    (a) Optical potential in the continuum, with the Amman-Beenker (AB) tiling overlaid in black. (b) Corresponding tight-binding Hamiltonian. Each dot represents the centre of mass of a numerically computed Wannier function, color-coded by onsite energy. Tunnelling amplitudes are shown in encoded in color of the line connecting sites. The lattice sites closely follow the AB tiling, and the onsite energies and tunneling amplitudes form intricate quasiperiodic patterns. The short diagonal of the rhombuses exhibit significant tunneling amplitudes, in stark contrast with the usual AB tiling where these bonds are absent. Positive tunneling amplitudes are shown but are significantly weaker than negative tunnelings. \textit{Parameters:} $V_0 = 1.5 \ E_r$.}
    \label{fig:8QC_real_space}
\end{figure}

To obtain a precise understanding of the physical properties of quasiperiodic systems, it is necessary to work with large system sizes in order to resolve fine spectral details.
In this work, we focus on the energy band spanned by the ground Wannier functions of the potential, and leave the treatment of higher bands to future studies. We use a large-scale tight-binding Hamiltonian of the lowest energy band of the 8QC constructed using an extension of our methods developed in \citep{gottlobHubbardModelsQuasicrystalline2023}. The lowest-band Wannier functions, except for well-understood exceptions \citep{gottlobHubbardModelsQuasicrystalline2023}, are  localized close to the local minima of the potential. While the original method in \citep{gottlobHubbardModelsQuasicrystalline2023} solved the Schrödinger equation on a fixed real-space finite-difference grid around each lattice site, we now parameterize the wavefunctions using the \textit{sinc-discrete variable representation} (sinc-DVR) \citep{wallEffectiveManybodyParameters2015, weiHubbardParametersProgrammable2024}. This change preserves the essential procedure of solving the Schrödinger equation numerically but uses a more efficient representation of the basis. The sinc-DVR substantially reduces the memory requirements for storing all individual Wannier functions $\ket{w_i}$ by lowering the number of required discretization points, thereby allowing us to increase the accessible system size from about $1600$ sites to around $12000$ sites; see Appendix \ref{app:sincdvr} for details.

We construct the tight-binding Hamiltonian by evaluating the matrix elements of the real-space continuum Hamiltonian $\hat{H}_{cont} = -\frac{\hbar^2}{2m}\Delta + V(x)$ in the Wannier basis for the on-site energies $\epsilon_i=\bra{w_{i}}\hat{H}_{cont} \ket{w_{i}}$ and  hopping amplitudes  $J_{ij} = \bra{w_{i}}\hat{H}_{cont} \ket{w_{j}}$. The Wannier basis can also be used to calculate the interaction energies, but these are irrelevant in the present context. The resulting tight-binding Hamiltonian
\begin{equation}
    \hat{H}_{8QC} = \sum_{i} \epsilon_i \hat{a}_i^\dagger \hat{a}_i + \sum_{i\neq j} J_{ij} \hat{a}_i^\dagger \hat{a}_j \nonumber 
\label{eq:8QCTB} 
\end{equation}

allows us to gain insight into the microscopic parameters governing the  physics of the 8QC, and to study its eigenstates and energy spectrum at large system size. A small section of the Hamiltonian is represented graphically in \cref{fig:8QC_real_space}, where each lattice sites is defined as the center of mass of the corresponding Wannier function. The resulting tight-binding Hamiltonian is composed of the lattice sites sitting in a circular patch of the 8QC of diameter $70\,\lambda$, and contains  around four times more lattice sites than is achievable by direct numerical diagonalisation of the continuum Hamiltonian \citep{ZhuLocalization2024}. We stress that the so-obtained tight-binding Hamiltonian is an \textit{exact} model of the 8QC lowest energy band, and that the Wannier functions form a complete basis of that lowest energy band, and as such capture the eigenstates and energy spectrum of the continuum Hamiltonian exactly. The numerical codes for constructing the Wannier functions are available in the following online repository \citep{gottlobEmmgottlobQuasiHubbard2024a}, and the large-scale tight-binding Hamiltonians are available at \citep{8QCHamiltonian}.

\begin{figure}
    \centering
    \includegraphics[width = \linewidth]{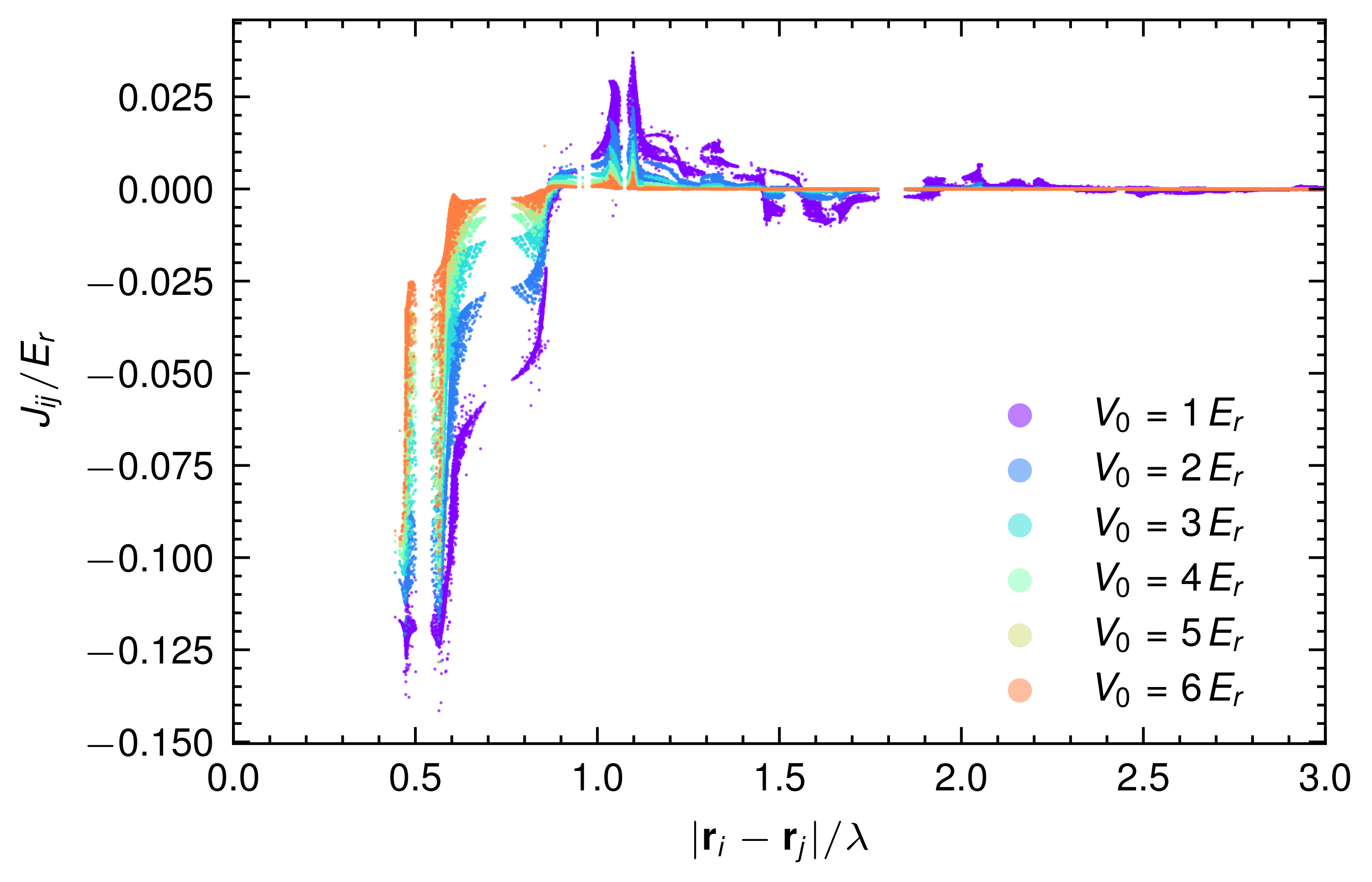}
    \caption{\textbf{Tunneling amplitudes of the 8QC as a function of distance between sites:} The tunneling amplitudes exhibit an oscillating decay as a function of distance, which is caused by the oscillating sidelobes of the Wannier functions, analogous to periodic lattices. The tunnelling range decreases rapidly with lattice depth, indicating a suppression of longer-range tunneling processes.}
    \label{fig:8QC_J_distance}
\end{figure}

A detailed analysis of the distributions of the onsite energies and tunneling amplitudes as a function of the lattice depth $V_0$ can be found in \citep{gottlobHubbardModelsQuasicrystalline2023}. Most relevant to this work are the fact that the width of the onsite energy distribution increases monotonously with the lattice depth $V_0$, and that the tunneling amplitudes decrease approximately exponentially with distance and decay strongly with $V_0$.
It is worth noting that the tunneling amplitudes oscillate as a function of the bond length $|\mathbf{r}_i - \mathbf{r}_j|$, due to the oscillating sidelobes of the Wannier functions (\cref{fig:8QC_J_distance}).

\section{Spectral and localisation properties} \label{sec:spectral_localisation}

\begin{figure*}
  \centering
  \includegraphics[width = 18cm]{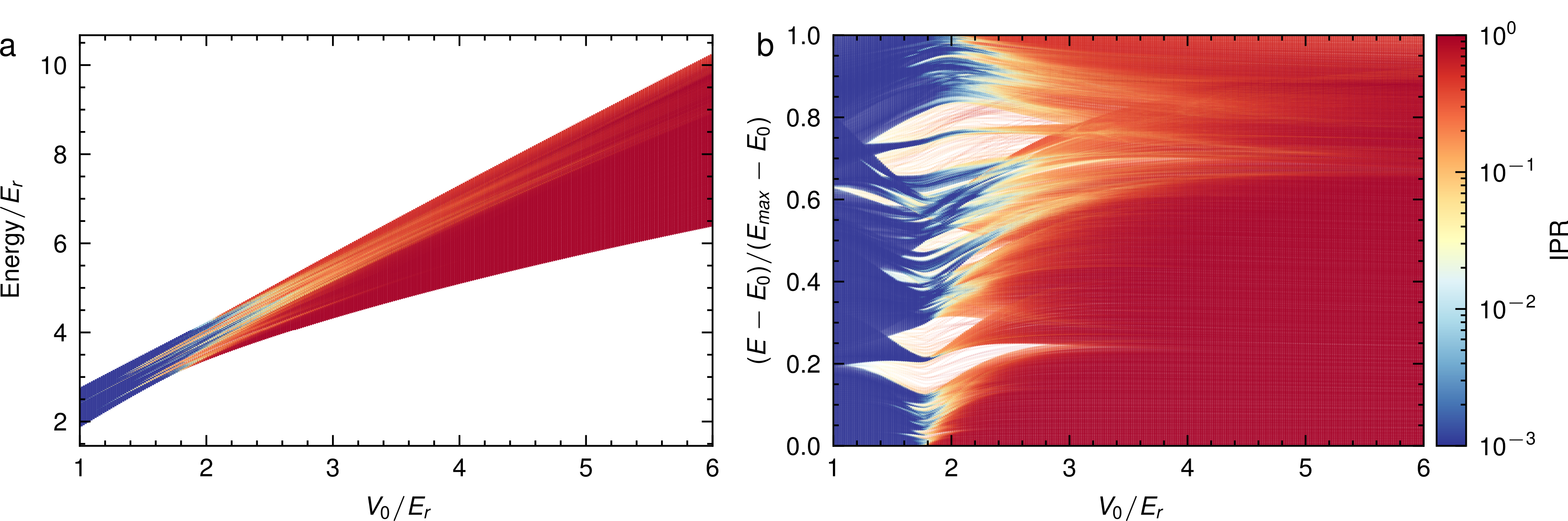}
  \caption{\textbf{Localisation properties of the non-interacting energy spectrum of the (lowest-band) of the 8QC:} Color encodes the IPR of the eigenstates. (\textbf{a}) Energy spectrum. (\textbf{b}) Energy spectrum normalized by bandwidth, where the energy of the ground-state has been subtracted. The spectrum contains a hierarchy of gaps (visible as white spaces). The ground state localizes at $1.76 \ E_r$, and the majority of the spectrum localizes between $1.76 \ E_r$ and around $3 \ E_r$. Faint in-gap lines correspond to edge states caused by the open boundary conditions. \textit{Parameters:} System diameter $70 \lambda$.}
  \label{fig:IPR}
\end{figure*}

Using the above described tight-binding Hamiltonian of the 8QC, we compute the single-particle energy spectrum and eigenstates of the large-scale lattice. At intermediate lattice depths, the energy spectrum (\cref{fig:IPR}) contains a series of minigaps, in good agreement with numerical solutions of the single-particle Schrödinger equation in the continuum~\citep{ZhuLocalization2024}.

 Spectral properties, such as level-spacing statistics, are often closely interlinked with the localisation properties of the eigenstates. 
In Anderson localized systems, for instance, localized eigenstates can be spatially separated by large distances. Hence they cannot hybridize and can therefore evade the level repulsion typical for ergodic systems and be almost degenerate \citep{ShklovskiiStat1993}. 

Therefore, to understand the origin of energy gaps in the quasicrystalline potential, we first turn to studying the localisation properties of its eigenstates $\psi_E(i)$ by computing their Inverse Participation Ratio (IPR) in the Wannier basis:
\begin{equation}
  \text{IPR}_n = \sum_i |\psi_n(i)|^4 \,,
\end{equation}
where the index $n$ runs over energies, and $i$ runs over lattice sites. \cref{fig:IPR} shows the IPR of all eigenstates. An IPR of 1 implies that the considered eigenstate is maximally localized and consists entirely of a single Wannier function, while the IPR of an extended state will vanish in the limit of infinite system size.
We find that the energy spectrum seems to ``pinch" (i.e.\ the measure of the energy spectrum decreases strongly) around the localisation transition, which is reminiscent of the zero measure of the energy spectrum of the Aubry-André-Harper chain at criticality \citep{aubry1980annals}. Unlike the Aubry-André chain however, the 8QC undergoes an energy-dependent localisation transition, with an energy dependence that does not seem smooth in energy. This is in agreement with a previous study which analyzed the scaling of the IPR (in the continuum real-space basis) with system size for the 8QC, and showed that the localisation transition did not have a smooth energy dependence \citep{ZhuLocalization2024}.

\begin{figure}
    \centering
    \includegraphics[width = 8.9cm]{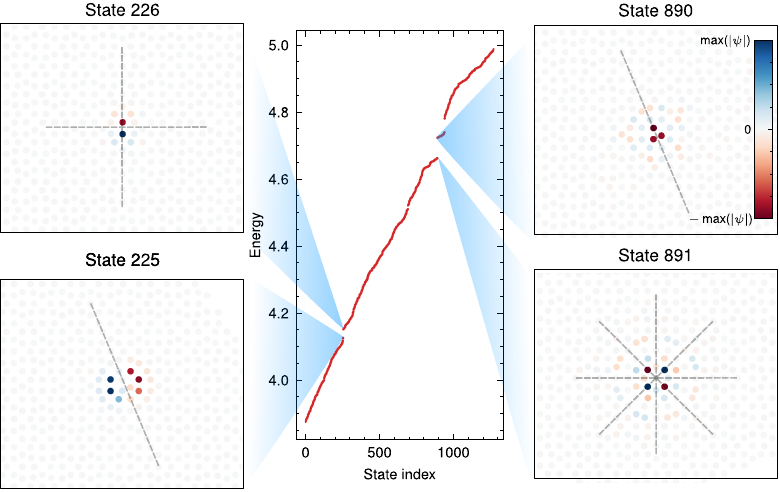}
    \caption{\textbf{Eigenstates around the largest energy gaps in the localized phase:} selection of pairs of states surrounding two of the largest gaps. Interestingly, these states are localized around local symmetry centres with various levels of symmetries (grey lines indicate approximate local reflection symmetries of the lattice) and are composed of small groups of resonant sites. Edge states have been filtered out by ignoring eigenstates with more than $2.5 \%$ of probability within $1 \ \lambda$ from the system's edge. All eigenstates are shown centered around their center-of-mass, they are situated in different regions of the 8QC. \textit{Parameters: } $V_0=2.5 \ E_r$. System diameter $25 \ \lambda$.}
    \label{fig:8QCgapstates}
\end{figure}

In \cref{fig:8QCgapstates}, we inspect the eigenstates that are located directly below and above the largest spectral gaps  at an intermediate lattice depth ($V_0=2.5 \ E_r$) well above the ground state localization transition. Interestingly, we find that around some energy gaps, the eigenstates are formed by hybridized collections of few (nearly-) resonant sites, which exhibit various levels of symmetries (reflection symmetry lines are indicated by gray lines). This suggests that the energy gaps in the 8QC are related to resonant sites arranged around approximate local symmetry centers (i.e.\ locations in the potential whose local surroundings approximately symmetric). This observation calls for a more systematic study of the mechanism behind the formation of energy gaps in the 8QC, and quasicrystalline potentials in general, which we address in \cref{sec:8QCspectrumconfig}.

\section{Infinite-size limit: configuration-space description} \label{sec:confighubbard}

To go beyond finite-size numerics, we use the configuration-space description of the 8QC, which we previously developed in \citep{gottlobHubbardModelsQuasicrystalline2023} to describe the 8QC in the infinite-size limit. Here, we recall the steps for constructing the configuration-space representation and its main features. 

\begin{figure}
  \centering
  \includegraphics[width = \linewidth]{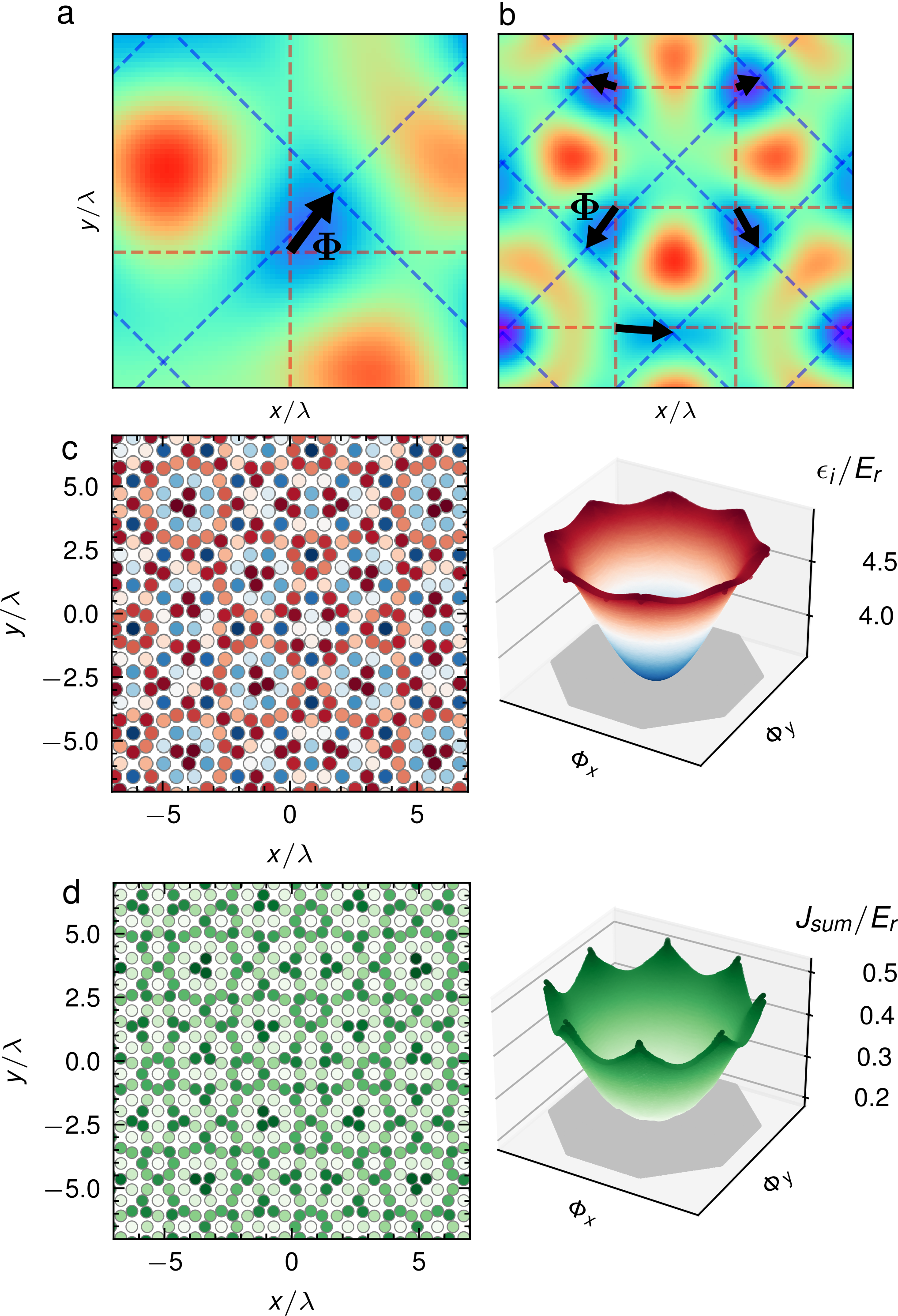}
  \caption{\textbf{Configuration-space representation of the 8QC}: (a,b) For every lattice site, the vector $\mathbf{\Phi}$ (black arrows) quantifies the local displacement between the closest minima of the XY (red dotted lines) and D (blue dotted lines) square lattices, and hence for each site determines its shape and local environment. (c,d) The tight-binding Hamiltonian (left) can  be re-expressed in configuration space (right),  In the infinite-size limit, this forms a uniformly dense octagon with periodic boundaries, where the onsite energies (c) and tunneling amplitudes (d) form smooth and 8-fold symmetric surfaces $\epsilon(\mathbf{\Phi})$ and $J(\mathbf{\Phi}, \mathbf{\Phi}')$. For display purposes, (d) shows the total tunneling amplitude per site $J_{sum}(\mathbf{\Phi_i}) = \sum_{i\neq j} |J_{ij}|$, instead of individual tunneling elements $J_{ij}$. \textit{Parameters:} System diameter $70 \, \lambda$. $V_0 = 2.5 \ E_r$.}
  \label{fig:BHoctagon}
\end{figure}

To obtain the configuration-space representation, we sort the lattice sites according to their local environment. To do so, we exploit the fact that the 8QC can be constructed by superimposing one square grid (formed by the $\mathbf{k}_x$, $\mathbf{k}_y$ beams, referred to as XY lattice) and one \textit{diagonal} square grid rotated by $45^\circ$ (formed by the $\mathbf{k}_+$, $\mathbf{k}_-$ beams, referred to as D lattice). For each lattice site  $\mathbf{r_i}$, we then define the local offset between the XY and the D lattice as
\begin{equation} \label{eq:configphi}
    \mathbf{\Phi}(\mathbf{r_i}) \equiv \mathbf{\Phi}_{XY}(\mathbf{r_i}) - \mathbf{\Phi}_{D}(\mathbf{r_i})
\end{equation} 
where the vectors $\mathbf{\Phi}_{XY}$ and $\mathbf{\Phi}_{D}$ encode the position of the lattice site within the unit cell of the XY and D lattice
\begin{align} \label{eq:phases_XY}
  \mathbf{\Phi}_{XY} (\mathbf{r}) &= \left[\left( x+\frac{\phi_1}{k}\right) \mathrm{mod}\; d \right] \, \,  {\mathbf{e}_x} \nonumber\\
              &+ \left[ \left( y+\frac{\phi_2}{k}\right) \mathrm{mod}\; d \right] \, \,   {\mathbf{e}_y}
\end{align}
\begin{align} \label{eq:phases_D}
    \mathbf{\Phi}_{D}(\mathbf{r}) &= \left[\left( \frac{x+y}{\sqrt{2}}+\frac{\phi_3}{k_0} \right) \mathrm{mod}\; d \right] \, \,  {\mathbf{e}_+} \nonumber \\
                & + \left[\left( \frac{x-y}{\sqrt{2}}+\frac{\phi_4}{k_0} \right) \mathrm{mod}\; d \right] \, \,  {\mathbf{e}_-} \,
\end{align}
Here, we defined $d=\lambda/2$ as the lattice constant of both square lattices, $\mathbf{e}_x$, $\mathbf{e}_y$ are the unit-vectors along the $x$ and $y$ directions,  $\mathbf{e}_\pm = \frac{\mathbf{e}_x\pm \mathbf{e}_y}{\sqrt{2}}$,  and the $\phi_i$ are the lattices' phases introduced in \cref{eq:QCpotential}. 
For every site $\mathbf{r_i}$, the vector $\mathbf{\Phi}(\mathbf{r_i})$ then quantifies the local displacement between the XY and D lattices, that is the displacement between their closest minima. The vector $\mathbf{\Phi}(\mathbf{r_i})$ thus not only determines the shape of the potential minimum at site $i$ but also the configuration of its local environment (see \cref{fig:BHoctagon} a,b for examples).

Mapping all sites $\mathbf{r_i}$ to their position $\mathbf{\Phi}(\mathbf{r_i})$ in configuration space, we obtain an octagonal parameter space (of inner diameter $d$) with the following properties:
\begin{enumerate}
    \item Every $\mathbf{\Phi}$ uniquely identifies a single lattice site.
    \item  In the infinite-size limit, the octagon is densely and uniformly populated with lattice sites, which is similar to the perpendicular spaces of octagonal discrete  quasiperiodic lattices \citep{mirzhalilovPerpendicularSpaceAccounting2020, oktelStrictlyLocalizedStates2021}.
    \item Every edge of the octagon can be identified with the opposing edge, equipping configuration space with periodic boundary conditions and hence the topology of a two-hole torus.
    \item Symmetry points or lines in configuration space correspond to symmetry points or lines of the quasicrystal. For instance, the center of configuration space corresponds to the global minimum and center of eight-fold symmetry of the 8QC potential.
\end{enumerate}

We can then re-express the tight-binding Hamiltonian of the 8QC in this uniformly dense octagonal configuration space
by simply mapping the real-space coordinates $\mathbf{r}_i$ of all sites of the 8QC to the corresponding  $\mathbf{\Phi}(\mathbf{r}_i)$:
\begin{align} 
\hat{H}_{8QC} &= \sum_{\mathbf{\Phi}} \epsilon(\mathbf{\Phi})  \hat{a}_{\mathbf{\Phi}}^\dagger \hat{a}_{\mathbf{\Phi}} \nonumber\\
& + \sum_{\mathbf{\Phi} \neq \mathbf{\Phi}'} J(\mathbf{\Phi}, \mathbf{\Phi}') \hat{a}_{\mathbf{\Phi}}^\dagger \hat{a}_{\mathbf{\Phi'}} 
\label{eq:BHconfig} \end{align}

Here, $\hat{a}_{\mathbf{\Phi}}$ is the annihilation operator for the Wannier function at the site with coordinate $\mathbf{\Phi}$ in configuration space.
This expression emphasizes that in configuration space -- which is a compact and uniformly dense space with periodic boundaries -- the 8QC is entirely described by the functions $\epsilon(\mathbf{\Phi})$ and $J(\mathbf{\Phi}, \mathbf{\Phi}')$, as shown in \cref{fig:BHoctagon}. Contrary to the fractal and quasi-disordered structure exhibited in real space, in configuration space the tight-binding Hamiltonian is composed of functions that are smooth. 

For lattice sites close to the centre of the octagonal configuration space ($|\mathbf{\Phi}|\ll d)$, the minima of the two square lattices are well overlapped such that these sites correspond to the lowest potential wells in real space. They   are characterized by the lowest onsite energies and the lowest summed tunneling amplitudes $J_{sum}(\mathbf{\Phi_i}) = \sum_j |J_{ij}|$. In contrast, sites at the edges of the octagonal configuration space correspond to higher-lying and shallower potential wells in real-space and are characterized by higher values of onsite energies and higher summed tunneling amplitudes.

\begin{figure}
  \centering
  \includegraphics[width =\linewidth]{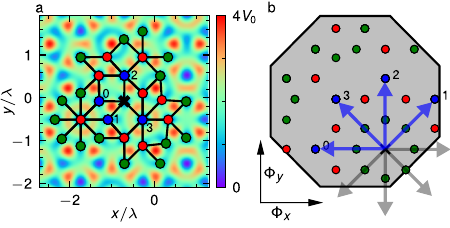}
  \caption{\textbf{Classification of neighbors in the 8QC:} 1st, 2nd and 3rd-order neighbors of a given site (black cross) in (a) real-  and (b) configuration-space constructed by $\mathbf{\Phi}' = \mathbf{\Phi}+\sum_i c_i \mathbf{\tilde{e}}_i$ with $c_i\in\mathbb{Z}$ and $\sum{|c_i|}=\{1,2,3\}$. (blue): 1st-order, (red): 2nd-order, (green): 3rd-order neighbors. Black lines connect 1st-order neighbors.}
  \label{fig:octagon_neighbors}
\end{figure}

Unlike in periodic systems, where nearest neighbors can be defined naturally using the lattice spacing, the definition of nearest neighbors in the 8QC is not straightforward. Nevertheless, configuration space can be used to derive a rigorous hierarchy of neighbors for the 8QC, as we showed in detail in \citep{gottlobHubbardModelsQuasicrystalline2023}.
In configuration space, nearest neighbors are defined as sites which are separated by the vectors
\begin{equation} \label{eq:etilde} \mathbf{\tilde{e}}_i \in \pm \frac{d}{1+\sqrt{2}} \left\{ \mathbf{e}_x, {\mathbf{e}_y}, {\mathbf{e}_+}, {\mathbf{e}_-} \right\}.
\end{equation}
As illustrated in \cref{fig:octagon_neighbors} for a chosen site (black cross) at position $\mathbf{\Phi}$, if $\mathbf{\Phi}+\mathbf{\tilde{e}}_i$ lies within the octagon, the corresponding site (blue) constitutes a close neighbor in real space, which is then defined as a \textit{first-order} neighbor. The same structure is found in the perpendicular space of eight-fold discrete quasicrystals \citep{oktelStrictlyLocalizedStates2021}. Note, due to the minus sign in front of $\mathbf{\Phi}_D$ in \cref{eq:configphi}, a move along the diagonal directions $\mathbf{e}_\pm$ in configuration space corresponds to a shift along $-\mathbf{e}_\pm$ in real-space.

In turn, this construction can be generalized to define the $n$th-order neighbors of a given site as the set of sites that lie within the octagon and are connected through the sum of at least $n$ vectors $\mathbf{\tilde{e}}_i$, i.e.,\ $\mathbf{\Phi}' = \mathbf{\Phi}+\sum_i c_i \mathbf{\tilde{e}}_i$ with $c_i\in\mathbb{Z}$ and $\sum{|c_i|}=n$. \cref{fig:octagon_neighbors} shows an example of 1st, 2nd and 3rd-order neighbors. With this definition, the edges (black lines) connecting 1st-order neighbors correspond to the edges of the square and rhombuses of the Ammann–Beenker tiling \citep{maceQuantumSimulation2D2016}. We note that strong tunneling amplitudes can connect pairs of sites along the short diagonals of the rhombuses, which are nevertheless 2nd-order neighbors.

\section{Localisation properties in real and configuration space} \label{sec:hybridisation}
To further characterize the localisation properties of the single-particle eigenstates of the 8QC, we compute the following quantity, which we refer to as the \textit{hybridisation ratio}:
\begin{equation}
    \text{HR}_{i} = \frac{1}{\sum_n |\psi_n(i)|^4} \,,
\end{equation}
where the index $n$ indexes the single-particle eigenstates, and $i$ refers to lattice sites. This quantity can be thought of as a \textit{site-resolved} participation ratio, in contrast to the traditional \textit{energy-resolved} participation ratio, and quantifies the number of eigenstates which are supported on a given site $i$. In other words, the HR measures the degree of hybridisation of a given site with all other lattice sites. For instance, a site hybridized with exactly one neighbor will be characterized by a HR of $2$.

\begin{figure*}
    \centering
    \includegraphics[width =\linewidth]{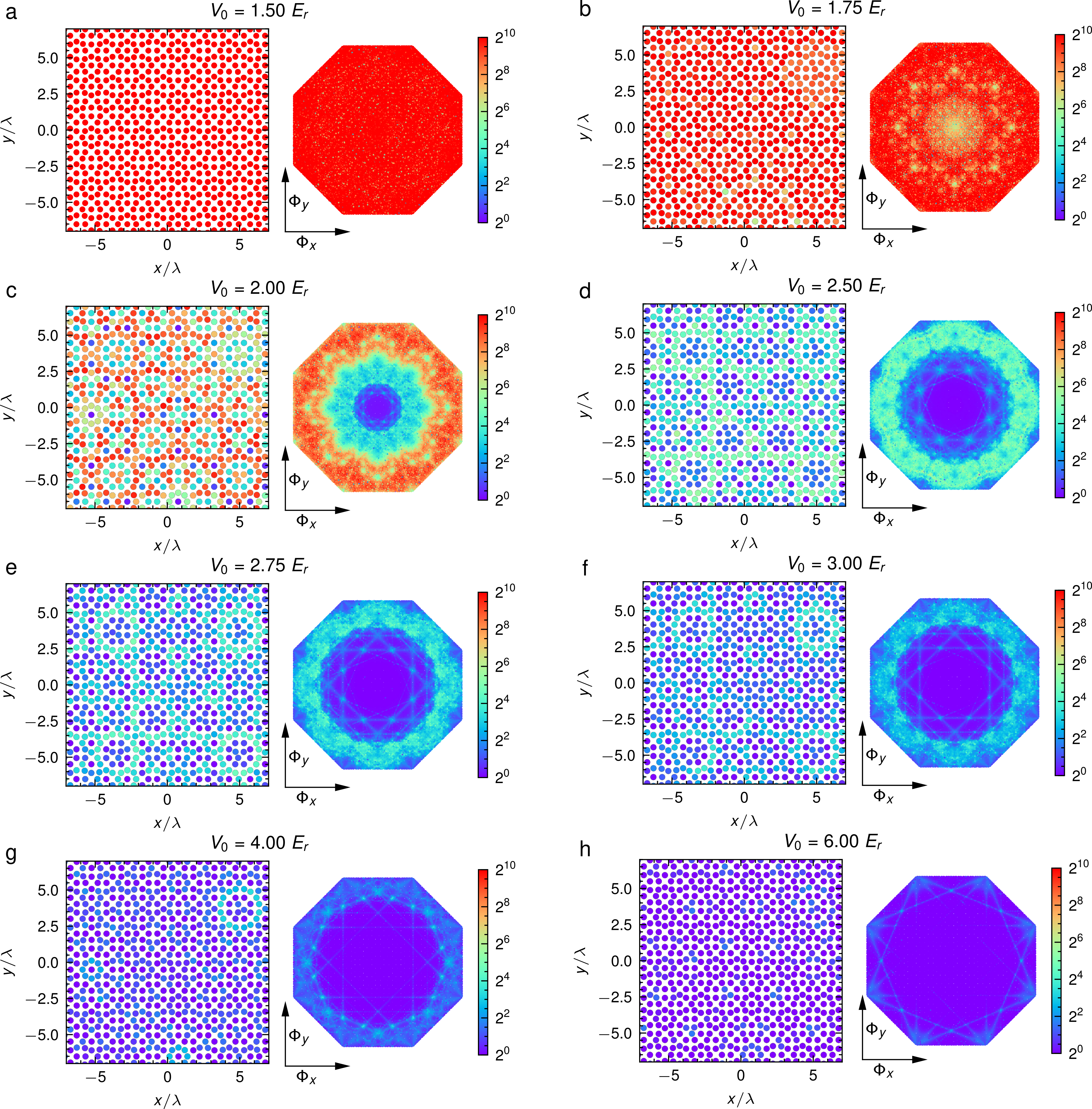}
    \caption{\textbf{Hybridisation ratios in the 8QC, in real- (left) and configuration- (right) space:} the hybridisation ratio quantifies with how many other sites a given site is hybridized. (\textbf{a}) Below the ground-state localisation transition, all sites have  high HR (note, the colorbar is cut off at $2^{10}$), consistent with extended eigenstates spanning the whole lattice. (\textbf{b}) Close to the ground-state localisation transition at $1.76 \ E_r$, the centre of the octagon starts to exhibit lower HR. (\textbf{c}) Above the ground-state localisation transition low-lying sites around the central portion of the octagon host strongly localized states and become decoupled from the rest of the lattice. (\textbf{d}, \textbf{e}, \textbf{f}) The lattice is split into a growing group of decoupled sites around the centre of configuration space, and shrinking connected networks of hybridized sites (outer regions of configuration space). For the deeper regimes (\textbf{g}, \textbf{h}), the entire lattice is strongly localized, showing only few sites that still hybridize with $1$ or $2$ resonant neighbors.  On the octagon, clear straight lines with higher HR are visible, which we explain as resonant lines in \cref{sec:hybridisation}. In the deep lattice depth regime (h), the hybridization is confined very sharply around the resonant lines and hence stops being consistently sampled by the finite system size used in the calculations. \textit{Parameters:} System diameter $70 \ \lambda$. }
    \label{fig:hybridisation_ratio}
\end{figure*}

\cref{fig:hybridisation_ratio} shows the HR for several lattice depths, below and above the ground-state localisation transition. Below the transition, all sites have a large HR, due to the delocalized nature of the entire spectrum. On approaching the ground-state localisation transition (\cref{fig:hybridisation_ratio} b), the HR of some sites starts to decrease. In particular, the first sites to host strongly localized states and thereby to ``decouple" from the rest of the lattice are at the center of configuration space, which we recall are characterized by the lowest onsite energies and lowest tunneling amplitudes. As the lattice depth increases, the fraction of decoupled sites increases, while other sites more towards the outer parts of configuration space still form a well-connected network of hybridized sites. The relatively sharp demarcations in configuration space  suggest that the lattice sites might be separated by a mobility edge in configuration space rather then in energy, reminiscent of the mixed spectra found in 2D Aubry-Andre models~\cite{szaboMixedSpectraPartially2020}. 

Finally, for even higher lattice depths (\cref{fig:hybridisation_ratio} h), the system becomes a collection of isolated sites that hybridize with at most $1$ or $2$ neighbors such that all eigenstates are strongly localized. This establishes that, in contrast to the 2D Aubry-Andre models~\cite{szaboMixedSpectraPartially2020},  the 8QC does not contain any weakly-modulated lines  that could potentially host delocalized eigenstates even at relatively high lattice depths. This finding, combined with the non-existence of rare regions in the 8QC, turns them into ideal systems to search for a two-dimensional  many-body localized phase.

\begin{figure}
    \centering
    \includegraphics[width = 5cm]{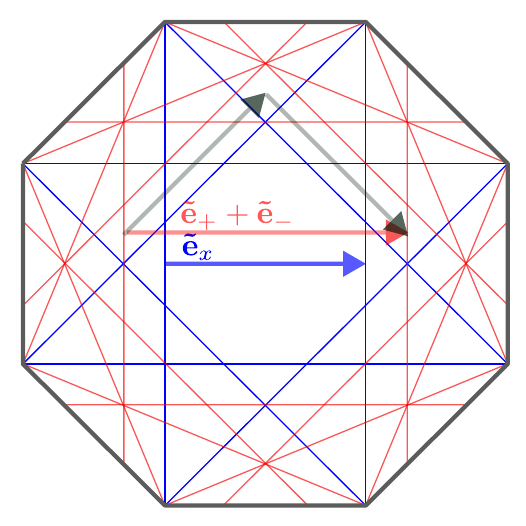}
    \caption{\textbf{Resonant lines in configuration space:} Due to the eight-fold symmetry of the onsite energies $\epsilon(\mathbf{\Phi})$ in configuration space, all sites lying on the blue lines will have a resonant first-order neighbor. For instance, the two vertical blue lines are separated by the vector $\mathbf{\tilde{e}}_x$, indicating that horizontal hopping connects every point on one line to a point on the other line. Due to the eight-fold symmetry, these two points have the same on-site energy and hence corresponds to resonant neighbors.
    Red lines indicate sites which have a resonant second-order neighbors.  Note, the figure only shows the subset of the second-order resonant lines that are associated with significant hybridisation.
    }
    \label{fig:resonantlines}
\end{figure}

Interestingly, within the localized region, the HR exhibits clear eight-fold symmetric sets of lines of sites with higher HR, which we explain as follows. We recall that first-order neighbors are separated by fixed vectors $\mathbf{\tilde{e}}_i$ in configuration space. In combination with the eight-fold symmetry of the onsite energies $\epsilon(\mathbf{\Phi})$ in configuration space, this results in pairs of specific \textit{resonant} lines in configuration space where every site on one line has a resonant real-space neighbor on the other line, see \cref{fig:resonantlines}. Since these pairs of sites can fully hybridize, lying on a resonant line results in a higher HR. The position of these resonant lines is governed purely by the geometry of the system and is hence independent of the precise functional form of $\epsilon(\mathbf{\Phi})$.  The lines seen in the numerically computed HR \cref{fig:hybridisation_ratio} (d,e,f) exactly match the first- and second-order resonant lines drawn on \cref{fig:resonantlines}; \cref{fig:hybridisation_ratio} (c) even shows higher-order lines at the center of configuration space. Note, the figure only shows the subset of the second-order resonant lines (in red) that can be associated with significant hybridsation. Additional second-order resonant lines that intersect close to the center of configuration space can be drawn, but do not correspond to strong hybridisation since tunneling elements are weaker for sites close the center of configuration space.

\section{Energy gaps due to local approximate symmetries}\label{sec:8QCspectrumconfig}
The structure of the 8QC's energy spectrum  is highly complex, see \cref{fig:IPR}, and due to the inapplicability of Bloch's theorem, requires dedicated theoretical tools to analyze it. In this section, we develop tools towards achieving a systematic description of the 8QC's energy spectrum, explaining the mechanisms responsible for the energy gaps and the associated bands.
In particular, we show how the resonant lines introduced in the previous section impact the spectral properties of the system. Note, we have previously applied the method we use in this section to derive the exact values of the integrated density of states associated with all the gaps of the one-dimensional Aubry-André-Harper chain \citep{gottlob2024quasiperiodicityprotectsquantizedtransport}.

We start by considering the deep lattice regime, where onsite energies vastly dominate over tunneling amplitudes. Ignoring tunneling in a first step, the eigenstates are all trivially localized on single lattice sites and the eigenenergies are given by the onsite energies, which are well captured by the analytical formula \citep{gottlobHubbardModelsQuasicrystalline2023}:
\begin{equation} \label{eq:8QCdeep}
    \epsilon(\mathbf{\Phi}) = \Delta_0 + \Delta(V_0) \sum_{i=1}^4 \text{sin}^2{(\mathbf{k}_i\cdot \mathbf{\Phi})}\,.
\end{equation}

From the uniform denseness of configuration space $\mathbf{\Phi}_i$ and the smoothness of $\cref{eq:8QCdeep}$ as a function of $\mathbf{\Phi}$, it follows that the energy spectrum is  dense in the thermodynamic limit in the absence of tunneling. This conclusion is corroborated numerically by noticing that the single-particle energy spectrum in the deep lattice regime (\cref{fig:IPR} b) does not exhibit any sizeable gap.

Introducing tunneling as a perturbation, one can use degenerate perturbation theory to infer that the 
largest effects of small tunneling appears whenever two neighboring sites are resonant, i.e., have the same on-site energy.
These resonant pairs form symmetric and anti-symmetric dimer states, split by an energy difference proportional to the tunneling amplitude. 
The sets of first- and second-order resonant neighbors lie on the resonant lines defined in the previous section, see \cref{fig:resonantlines}. The same argument can also be extended to the sets of $n$-th order neighbors. Along these lines, the resonant hybridisation induces a local energy splitting.

Importantly, the intersections of resonant lines define enclosed areas in configuration space. As soon as the local splitting \textit{along the edges} of a given enclosed area exceeds the variation of onsite energies along the edges, a global energy gap opens in the energy spectrum. The integrated density of states (IDoS), that is the number of eigenstates below a given energy $E$, at the resulting energy gap is then fixed by the area of the enclosed region in configuration space. While this reasoning rests on treating tunneling as a small perturbation, we see numerically that all energies vary smoothly with lattice depth.
Hence, the integrated densities of states at all gaps cannot vary with lattice depth. Therefore, this perturbative reasoning allows us to deduce the IDoS at these energy gaps in the thermodynamic limit and for all lattice depths.

\begin{figure}
     \centering
     \includegraphics[width=\linewidth]{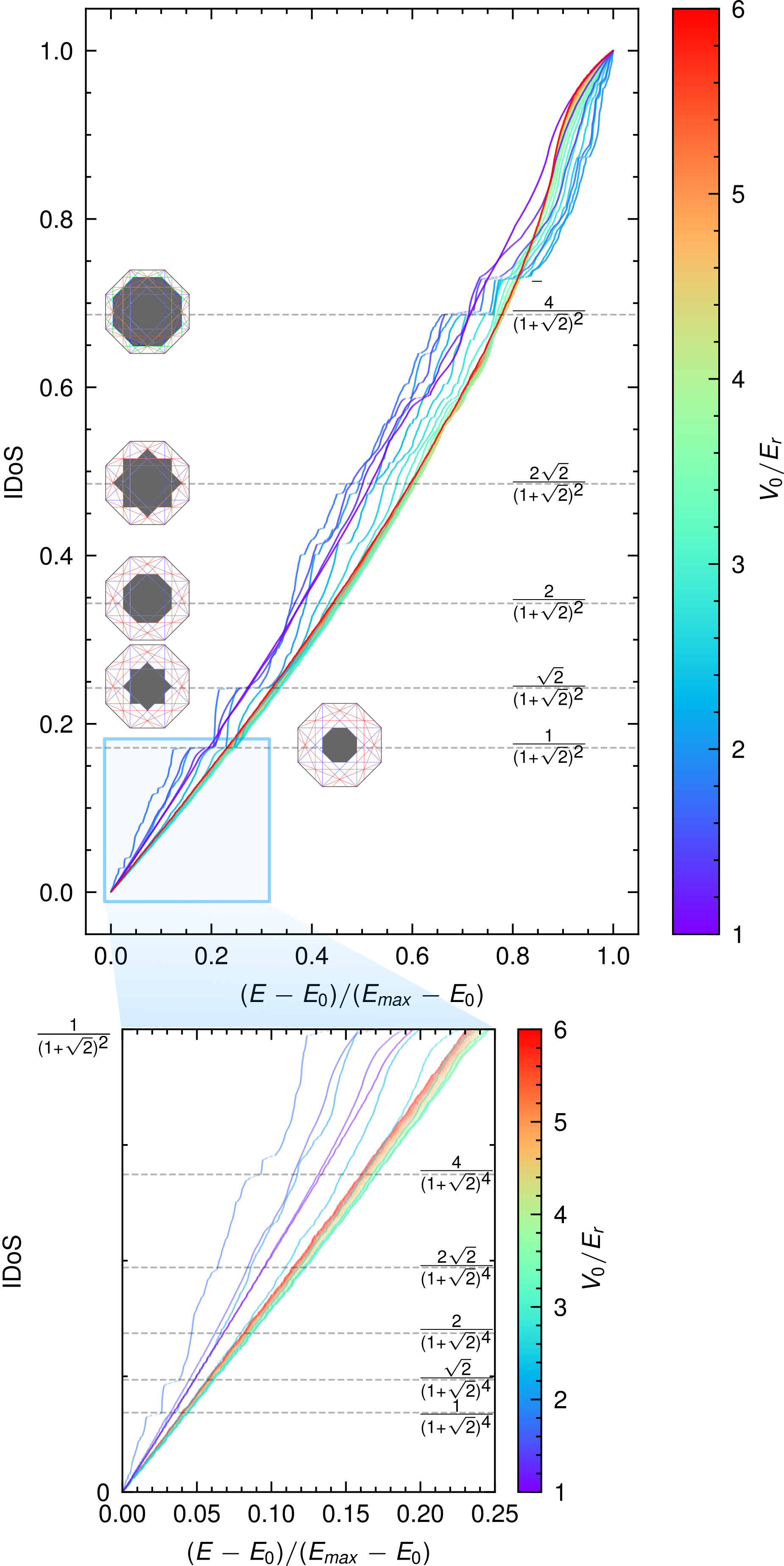}
     \caption{\textbf{Energy gaps in the 8QC from configuration-space arguments:} Integrated density of states (IDoS) of the non-interacting energy spectrum for various lattice depths $V_0$. We find that areas delimited by resonant lines in configuration space (grey areas in insets) can correspond to gaps in the energy spectrum as indicated by plateaus in the IDoS. 
     Inset: the lower part of the spectrum contains gaps whose IDoS reproduces the gaps in the main figure, exactly scaled down by a factor $\frac{1}{(\sqrt{2}+1)^2}$, and are therefore associated with third-order resonant lines. Insets: blue lines indicate first-order resonant lines, red lines are second-order and green lines are third-order resonant lines. \textit{Parameters:} System diameter of $70 \lambda$. }
     \label{fig:octagonareas}
 \end{figure}

We apply this reasoning to compute the integrated density of states below the lowest of the $4$ most significant gaps in the spectrum. This gap is caused by the resonant hybridisation occurring along the edges of the octagon of inner diameter $\frac{d}{1+\sqrt{2}}$ defined by the intersection of the first-order resonances, see lowest inset in \cref{fig:octagonareas}. As a consequence, the fraction of states contained below this gap must correspond to the ratio of the area of this octagon $A$ to the area of the whole configuration space $A_{tot}$:
\begin{equation} \label{eq:octagonfirstorder}
    \frac{A}{A_{tot}} = \frac{1}{(1+\sqrt{2})^2} \approx 0.1716
\end{equation}

 This is in excellent agreement with the value of $0.1717$ calculated numerically using our large-scale tight-binding model. We also apply the same reasoning to several other areas in configuration space (\cref{fig:octagonareas}), revealing that some of them do indeed match energy gaps very closely, even in the delocalized regime $V_0<1.76 \ E_r$ where the perturbative argument would a priori not be applicable. This offers a numerical confirmation that (at least some) energy gaps of the 8QC can be understood as originating from the hybridisation of resonant neighbors. It is however worth noting that, while for all the other areas depicted in \cref{fig:octagonareas}, the corresponding IDoSs match open gaps, the IDoS at $\frac{2}{(1+\sqrt{2})^2}$ does not correspond to an open gap in the spectrum.

In the localized regime of deeper lattices, the \textit{bandwidth} of the considered tight-binding model is dominated by the variation in onsite energies, while the above argument shows that the tunneling amplitudes primarily set the \textit{gap width} of the energy gaps in the localized phase of quasicrystalline potentials.

Further, we notice that the IDoS below the gap at $\frac{1}{(1+\sqrt{2})^2}$ exhibits a self-similar structure which perfectly repeats the sequence of gaps we have labelled above, with values of the IDoS that are simply scaled down by a factor $\frac{1}{(1+\sqrt{2})^2}$. Following our argument, we understand that these smaller gaps are caused by higher-order resonant neighbors; in this case by 3-rd order neighbors. Extending this reasoning \textit{ad infinitum} to the IDoS below $\frac{1}{(1+\sqrt{2})^4}$, we therefore expect that, provided that the corresponding tunneling amplitudes are strong enough, the lower part of the energy spectrum can potentially contain an infinite hierarchy of gaps, with each order of the hierarchy being related to the previous one by a factor of $\frac{1}{(1+\sqrt{2})^2}$.
 
\section{Energy bands composition}

To further analyze how the different energy bands are distributed in real and configuration space, we consider the \textit{windowed local density of states} (WLDoS)—i.e., the local density of states (LDoS) integrated over a finite energy interval $[E_{\min},E_{\max}]$. Formally, for site $i$,
\begin{equation}
\mathrm{WLDoS}(i;E_{\min},E_{\max}) \equiv\sum_{E\in [E_{\min},E_{\max}]} \bigl|\psi_E(i)\bigr|^2,
\label{eq:WLDoS}
\end{equation}
where $\psi_E(i)$ are the normalized eigenstates at energy $E$. 

\begin{figure*}
    \centering
    \includegraphics[scale = 0.76]{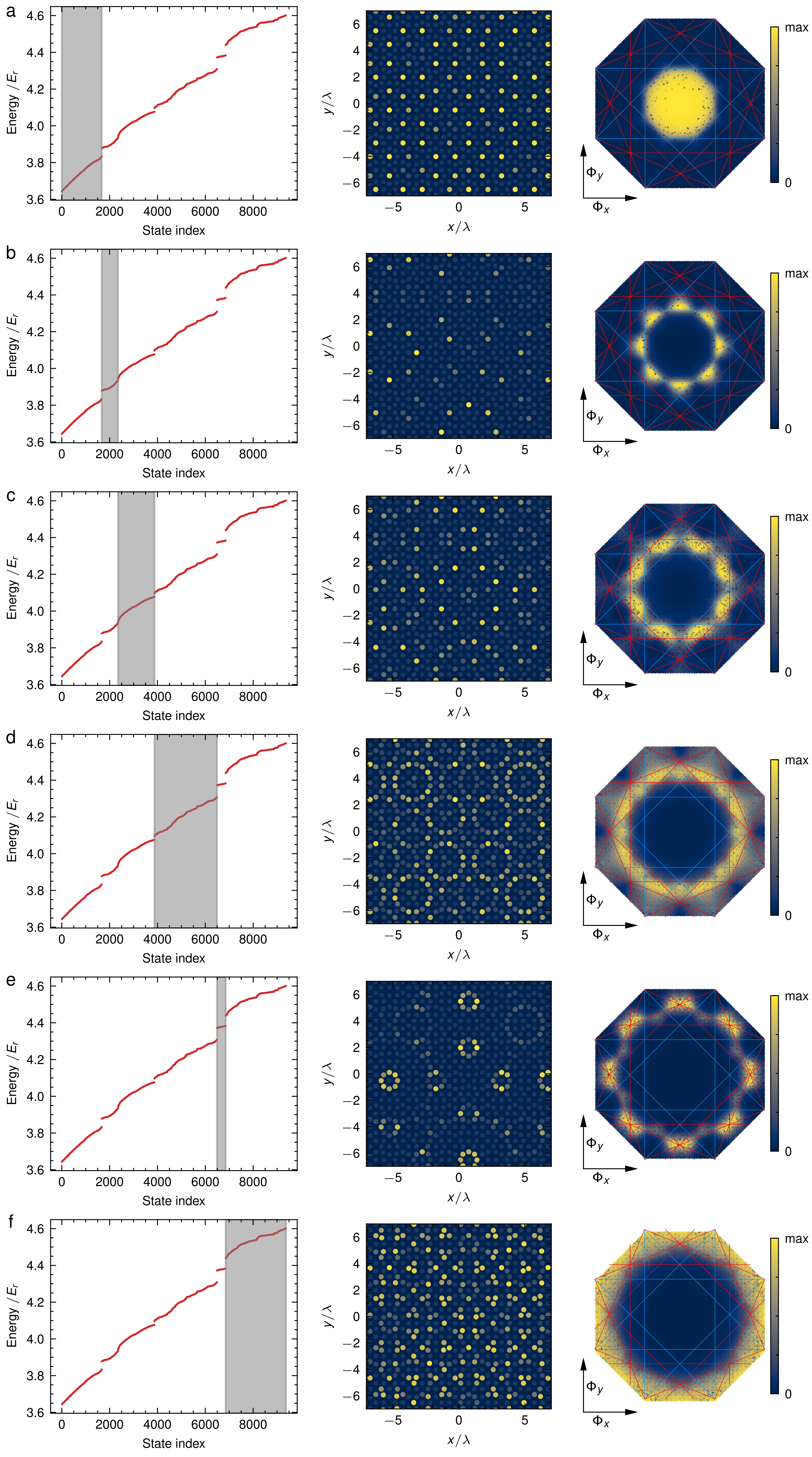}
    \caption{\textbf{Windowed Local Density of States in real  and configuration space in the localized phase:} Left: integration window. Middle: real-space WLDoS. Right: Configuration-space WLDoS. Blue (red) lines are first- (second-) order resonant lines, as defined previously.  Edge states have been filtered out by ignoring eigenstates with more than $2.5 \%$ of probability within $1 \, \lambda$ from the system's edge.  \textit{Parameters:} $V_0=2.25 \, E_r$. System diameter $70 \, \lambda$.}
    \label{fig:WLDoS}
\end{figure*}

 \cref{fig:WLDoS} show the WLDoS for several of the bulk energy bands surrounded by the largest energy gaps, where we filtered out the eigenstates localized at the edges of the system to better observe the bulk energy gaps. The left-hand column shows the energy window, the middle column shows the resulting integrated real-space local density for this energy window $\text{WLDoS}(i)$, and the right-hand column shows the corresponding density in configuration space.

While in real space it is difficult to characterize any structure in the WLDoS, the configuration-space picture exhibits vastly more interpretable patterns, whose structure does indeed closely follow regions defined by the intersection of resonant lines.  The higher-lying bands also occupy well-defined areas in configuration space, with the highest band mostly concentrated around the edges of the octagon. Interestingly, we observe in \cref{fig:WLDoS} (e) an almost flat band composed of eigenstates that seem to form localized states on approximately eight-fold symmetric rings containing eight sites. In configuration space, these sites lie at the intersection of second order resonant lines. The hybridisation between these 8 approximately resonant sites in turn creates two almost degenerate eigenmodes located at the average onsite energies of these $8$-site rings.
 
\section{Discussion and Conclusion} \label{sec:conclusion}

Quasicrystalline potentials are a platform of choice for studying the quantum properties of quasicrystals in the absence of lattice defects and lattice vibrations. Central to their many-body properties are the properties of their single-particle energy levels and eigenstates. As is often the case in the investigation of quasiperiodic systems, most of the previous studies were realized with finite-size systems, and very few analytical results  were known that apply in the thermodynamic limit. In particular, the existence and physical origin of true energy gaps (as opposed to pseudo-gaps, where the density of states is reduced but non-zero) in the energy spectrum of quasicrystalline potentials  was missing.  Of particular experimental relevance is the eight-fold quasicrystalline potential (8QC), which has been realized with ultracold atoms \citep{viebahnMatterWaveDiffractionQuasicrystalline2019,sbrosciaObservingLocalization2D2020} and was used to observe the existence of the elusive Bose glass phase \citep{Yu2024}.

In this work, we used the recently developed tool of configuration space -- which allows to describe the 8QC in the infinite-size limit -- to analytically elucidate the origin of energy gaps as resulting from energy splitting around approximate local symmetry centers and established that these gaps stay open in the thermodynamic limit.
Further, we showed that in the infinite-size limit the integrated densities of states below some of the system's energy gaps correspond to well-defined areas in configuration space, which take irrational values. 
To verify our results numerically, we built a large-scale numerical tight-binding Hamiltonian of the 8QC by extending the methods developed in \citep{gottlobHubbardModelsQuasicrystalline2023}, and observed excellent agreement with our analytical predictions.

First, we turned to the numerical study the tight-binding Hamiltonian of the 8QC. The energy spectrum contains a large number of minigaps, typical of quasiperiodic systems. We then analyzed the  hybridisation ratios of all lattice sites to reveal that the system undergoes a heterogeneous localisation transition, whereby different sites localize at different depths. Instead of being characterized by a mobility edge that separates localized from extended states in energy, it appeared that the localisation transition is approximately characterized by a mobility edge in configuration space. The localized fraction of the sites grows from the center of configuration space (where sites have lowest onsite energies) outwards, and the system eventually localizes entirely, establishing the absence of weakly modulated lines in the 8QC. Given that the 8QC does not contain Griffiths regions \citep{Griffiths1969, Agarwal2017}, the absence of weakly modulated lines in the deep lattice regime ($V_0 > 6 \ E_r$) established the 8QC as an ideal potential candidate for hosting a stable MBL phase in two dimensions.

Finally, we used the configuration-space representation to draw arguments about the system's energy gaps. In particular, we found the existence of a hierarchy of energy gaps resulting from level splitting caused by the resonant hybridisation of neighboring lattice sites. In turn, we showed analytically that the integrated densities associated with these gaps were given by specific (irrational) areas in configuration space, related to powers the silver ratio $\frac{1}{1+\sqrt{2}}$. To demonstrate the validity of the analytical construction, we predicted the integrated density of states expected below the lowest of the most significant gaps in the system and found excellent agreement with numerics obtained on a large-scale tight-binding Hamiltonian constructed using localized Wannier functions. Additionally, we observed that the system forms an almost flat band in which eigenstates are localized on eight-fold symmetric rings composed of $8$ sites. Our analysis revealed that these sites lie at specific points in configuration space, where several resonant lines intersect, and we therefore concluded that the almost flat band emerged from the two degenerate eigenmodes of these rings composed of 8 resonant sites.

In future work, it would be natural to explore the extension of our perturbative argument to predict the opening of all energy gaps of the 8QC systematically. Going further, we hope that such a development could also provide a way of labeling the energy gaps with their respective topological invariant (which was achieved recently for discrete octagonal quasicrystals in \citep{Jagannathan2023}), and therefore a rigorous gap labeling theorem for the 8QC. Additionaly, it would be interesting to use configuration space to study adiabatic pumping and quantized currents in optical quasicrystals, extending the work done in the context of the one-dimensional Aubry-André-Harper chain \citep{gottlob2024quasiperiodicityprotectsquantizedtransport}.

In conclusion, our study has paved the way for exciting new research directions in the field of quasicrystals. The discovery of genuine energy gaps with associated irrational density of states prompts intriguing questions about the potential existence of insulating phases at irrational fillings, such as fermionic band insulators and Mott insulators. Furthermore, our work has demonstrated the effectiveness of the configuration-space representation as an analytical tool for investigating infinite-size quasicrystals, bringing our understanding of optical quasicrystals closer to that of their periodic counterparts.  These findings not only contribute to our fundamental understanding of quasicrystals but also open up new possibilities for future investigations and applications.

\begin{acknowledgments}

The authors acknowledges support  by the European Commission ERC Starting Grant QUASICRYSTAL, the
EPSRC Programme Grant QQQS (No. EP/Y01510X/1) and the Deutsche Forschungsgemeinschaft (DFG, German Research Foundation) via Research Unit FOR 2414 under project number 277974659. E.G. acknowledges support from the Cambridge Trust. We are thankful to Shaurya Bhave, Georgia Nixon, Chelsea Ou, Leanne Reeve, Bo Song, Michael Wu, Jr-Chiun Yu and Dan S. Borgnia for helpful discussions.

\end{acknowledgments}

\section*{Data availability}
The numerical codes for constructing the Wannier functions are available in the following online repository \citep{gottlobEmmgottlobQuasiHubbard2024a}, and the large-scale tight-binding Hamiltonians are available at \citep{8QCHamiltonian}.

\bibliography{apssamp}

@article{Crowley2022,
  title = {Mean-field theory of failed thermalizing avalanches},
  author = {Crowley, P. J. D. and Chandran, A.},
  journal = {Phys. Rev. B},
  volume = {106},
  issue = {18},
  pages = {184208},
  numpages = {13},
  year = {2022},
  month = {Nov},
  publisher = {American Physical Society},
  doi = {10.1103/PhysRevB.106.184208},
  url = {https://link.aps.org/doi/10.1103/PhysRevB.106.184208}
}

@article{Vardeny2013,
  title = {Optics of photonic quasicrystals},
  volume = {7},
  ISSN = {1749-4893},
  url = {http://dx.doi.org/10.1038/nphoton.2012.343},
  DOI = {10.1038/nphoton.2012.343},
  number = {3},
  journal = {Nature Photonics},
  publisher = {Springer Science and Business Media LLC},
  author = {Vardeny,  Z. Valy and Nahata,  Ajay and Agrawal,  Amit},
  year = {2013},
  month = feb,
  pages = {177–187}
}

@article{Steurer2007,
  title = {Photonic and phononic quasicrystals},
  volume = {40},
  ISSN = {1361-6463},
  url = {http://dx.doi.org/10.1088/0022-3727/40/13/R01},
  DOI = {10.1088/0022-3727/40/13/r01},
  number = {13},
  journal = {Journal of Physics D: Applied Physics},
  publisher = {IOP Publishing},
  author = {Steurer,  Walter and Sutter-Widmer,  Daniel},
  year = {2007},
  month = jun,
  pages = {R229–R247}
}

@article{agrawalQuasiperiodicManybodyLocalization2022,
  title = {Quasiperiodic Many-Body Localization Transition in Dimension \$d{$>$}1\$},
  author = {Agrawal, Utkarsh and Vasseur, Romain and Gopalakrishnan, Sarang},
  year = {2022},
  month = sep,
  journal = {Physical Review B},
  volume = {106},
  number = {9},
  pages = {094206},
  doi = {10.1103/PhysRevB.106.094206},
  urldate = {2023-01-11}
}

@article{agrosiNaturallyOccurringAlCuFeSi2024,
  title = {A Naturally Occurring {{Al-Cu-Fe-Si}} Quasicrystal in a Micrometeorite from Southern {{Italy}}},
  author = {Agros{\`i}, Giovanna and Manzari, Paola and Mele, Daniela and Tempesta, Gioacchino and Rizzo, Floriana and Catelani, Tiziano and Bindi, Luca},
  year = {2024},
  month = feb,
  journal = {Communications Earth \& Environment},
  volume = {5},
  number = {1},
  pages = {1--6},
  issn = {2662-4435},
  doi = {10.1038/s43247-024-01233-w},
  urldate = {2024-05-06},
  copyright = {2024 The Author(s)},
  langid = {english}
}

@article{Yu2024,
  title = {Observing the two-dimensional Bose glass in an optical quasicrystal},
  volume = {633},
  ISSN = {1476-4687},
  url = {http://dx.doi.org/10.1038/s41586-024-07875-2},
  DOI = {10.1038/s41586-024-07875-2},
  number = {8029},
  journal = {Nature},
  publisher = {Springer Science and Business Media LLC},
  author = {Yu,  Jr-Chiun and Bhave,  Shaurya and Reeve,  Lee and Song,  Bo and Schneider,  Ulrich},
  year = {2024},
  month = sep,
  pages = {338–343}
}

@article{ahnDiracElectronsDodecagonal2018,
  title = {Dirac Electrons in a Dodecagonal Graphene Quasicrystal},
  author = {Ahn, Sung Joon and Moon, Pilkyung and Kim, Tae-Hoon and Kim, Hyun-Woo and Shin, Ha-Chul and Kim, Eun Hye and Cha, Hyun Woo and Kahng, Se-Jong and Kim, Philip and Koshino, Mikito and Son, Young-Woo and Yang, Cheol-Woong and Ahn, Joung Real},
  year = {2018},
  month = aug,
  journal = {Science},
  volume = {361},
  number = {6404},
  pages = {782--786},
  issn = {0036-8075, 1095-9203},
  doi = {10.1126/science.aar8412},
  urldate = {2020-10-09},
  chapter = {Report},
  copyright = {Copyright {\copyright} 2018 The Authors, some rights reserved; exclusive licensee American Association for the Advancement of Science. No claim to original U.S. Government Works. http://www.sciencemag.org.ezp.lib.cam.ac.uk/about/science-licenses-journal-article-reuseThis is an article distributed under the terms of the Science Journals Default License.},
  langid = {english},
  pmid = {29954987}
}

@article{aubry1980annals,
  title = {Annals Israel Phys},
  author = {Aubry, S and Andr{\'e}, G},
  year = {1980},
  journal = {Soc.},
  volume = {3},
  pages = {133}
}

@article{bellissardGapLabellingTheorems1992,
  title = {Gap Labelling Theorems for One Dimensional Discrete Schr{\"o}dinger Operators},
  author = {Bellissard, J. and Bovier, Anton and Ghez, Jean-Michel},
  year = {1992},
  journal = {Reviews in Mathematical Physics},
  volume = {04},
  number = {01},
  eprint = {https://doi.org/10.1142/S0129055X92000029},
  pages = {1--37},
  doi = {10.1142/S0129055X92000029}
}

@article{bindiAccidentalSynthesisPreviously2021,
  title = {Accidental Synthesis of a Previously Unknown Quasicrystal in the First Atomic Bomb Test},
  author = {Bindi, Luca and Kolb, William and Eby, G. Nelson and Asimow, Paul D. and Wallace, Terry C. and Steinhardt, Paul J.},
  year = {2021},
  month = may,
  journal = {Proceedings of the National Academy of Sciences},
  volume = {118},
  number = {22},
  issn = {1091-6490},
  doi = {10.1073/pnas.2101350118}
}

@article{bindiElectricalDischargeTriggers2023,
  title = {Electrical Discharge Triggers Quasicrystal Formation in an Eolian Dune},
  author = {Bindi, Luca and Pasek, Matthew A and Ma, Chi and Hu, Jinping and Cheng, Guangming and Yao, Nan and Asimow, Paul D and Steinhardt, Paul J},
  year = {2023},
  journal = {Proceedings of the National Academy of Sciences},
  volume = {120},
  number = {1},
  pages = {e2215484119}
}

@article{bindiEvidenceExtraterrestrialOrigin2012a,
  title = {Evidence for the Extraterrestrial Origin of a Natural Quasicrystal},
  author = {Bindi, Luca and Eiler, John M. and Guan, Yunbin and Hollister, Lincoln S. and MacPherson, Glenn and Steinhardt, Paul J. and Yao, Nan},
  year = {2012},
  month = jan,
  journal = {Proceedings of the National Academy of Sciences},
  volume = {109},
  number = {5},
  pages = {1396--1401},
  issn = {0027-8424, 1091-6490},
  doi = {10.1073/pnas.1111115109},
  urldate = {2024-05-06},
  langid = {english}
}

@article{ciardiFinitetemperaturePhasesTrapped2022,
  title = {Finite-Temperature Phases of Trapped Bosons in a Two-Dimensional Quasiperiodic Potential},
  author = {Ciardi, Matteo and Macr{\`i}, Tommaso and Cinti, Fabio},
  year = {2022},
  month = jan,
  journal = {Physical Review A},
  volume = {105},
  number = {1},
  pages = {L011301},
  doi = {10.1103/PhysRevA.105.L011301},
  urldate = {2022-11-01}
}

@article{deroeckStabilityInstabilityDelocalization2017,
  title = {Stability and Instability towards Delocalization in Many-Body Localization Systems},
  author = {De Roeck, Wojciech and Huveneers, Fran{\c c}ois},
  year = {2017},
  month = apr,
  journal = {Physical Review B},
  volume = {95},
  number = {15},
  pages = {155129},
  issn = {2469-9950, 2469-9969},
  doi = {10.1103/PhysRevB.95.155129},
  urldate = {2020-05-01},
  langid = {english}
}

@article{gautierStronglyInteractingBosons2021,
  title = {Strongly {{Interacting Bosons}} in a {{Two-Dimensional Quasicrystal Lattice}}},
  author = {Gautier, Ronan and Yao, Hepeng and {Sanchez-Palencia}, Laurent},
  year = {2021},
  month = mar,
  journal = {Physical Review Letters},
  volume = {126},
  number = {11},
  pages = {110401},
  issn = {0031-9007, 1079-7114},
  doi = {10.1103/PhysRevLett.126.110401},
  urldate = {2021-06-09},
  langid = {english}
}

@misc{gottlobEmmgottlobQuasiHubbard2024a,
  title = {Emmgottlob/{{QuasiHubbard}}},
  author = {Gottlob, Emmanuel},
  year = {2024},
  month = may,
  urldate = {2024-06-18},
  copyright = {MIT}, 
   url = {https://github.com/emmgottlob/QuasiHubbard}
}

@article{gottlobHubbardModelsQuasicrystalline2023,
  title = {Hubbard Models for Quasicrystalline Potentials},
  author = {Gottlob, E. and Schneider, U.},
  year = {2023},
  month = apr,
  journal = {Physical Review B},
  volume = {107},
  number = {14},
  pages = {144202},
  doi = {10.1103/PhysRevB.107.144202},
  urldate = {2023-11-26}
}

@article{kraus2012topological,
  title = {Topological States and Adiabatic Pumping in Quasicrystals},
  author = {Kraus, Yaacov E. and Lahini, Yoav and Ringel, Zohar and Verbin, Mor and Zilberberg, Oded},
  journal = {Physical Review Letters},
  volume = {109},
  issue = {10},
  pages = {106402},
  numpages = {5},
  year = {2012},
  month = {Sep},
  publisher = {American Physical Society},
  doi = {10.1103/PhysRevLett.109.106402},
  url = {https://link.aps.org/doi/10.1103/PhysRevLett.109.106402}
}

@article{Goblot2020,
  title = {Emergence of criticality through a cascade of delocalization transitions in quasiperiodic chains},
  volume = {16},
  ISSN = {1745-2481},
  url = {http://dx.doi.org/10.1038/s41567-020-0908-7},
  DOI = {10.1038/s41567-020-0908-7},
  number = {8},
  journal = {Nature Physics},
  publisher = {Springer Science and Business Media LLC},
  author = {Goblot,  V. and Štrkalj,  A. and Pernet,  N. and Lado,  J. L. and Dorow,  C. and Lemaître,  A. and Le Gratiet,  L. and Harouri,  A. and Sagnes,  I. and Ravets,  S. and Amo,  A. and Bloch,  J. and Zilberberg,  O.},
  year = {2020},
  month = jun,
  pages = {832–836}
}

@article{Ozawa2019,
  title = {Topological photonics},
  author = {Ozawa, Tomoki and Price, Hannah M. and Amo, Alberto and Goldman, Nathan and Hafezi, Mohammad and Lu, Ling and Rechtsman, Mikael C. and Schuster, David and Simon, Jonathan and Zilberberg, Oded and Carusotto, Iacopo},
  journal = {Rev. Mod. Phys.},
  volume = {91},
  issue = {1},
  pages = {015006},
  numpages = {76},
  year = {2019},
  month = {Mar},
  publisher = {American Physical Society},
  doi = {10.1103/RevModPhys.91.015006},
  url = {https://link.aps.org/doi/10.1103/RevModPhys.91.015006}
}

@book{Senechal1995,
  author    = {Marjorie Senechal},
  title     = {Quasicrystals and Geometry},
  year      = {1995},
  publisher = {Cambridge University Press},
  address   = {Cambridge}
}

@article{jagannathanEightfoldOpticalQuasicrystal2013,
  title = {An Eightfold Optical Quasicrystal with Cold Atoms},
  author = {Jagannathan, Anuradha and Duneau, Michel},
  year = {2013},
  month = dec,
  journal = {EPL (Europhysics Letters)},
  volume = {104},
  number = {6},
  pages = {66003},
  issn = {0295-5075, 1286-4854},
  doi = {10.1209/0295-5075/104/66003},
  urldate = {2020-04-14},
  langid = {english}
}

@article{duncanCritical2024,
  title = {Critical states and anomalous mobility edges in two-dimensional diagonal quasicrystals},
  author = {Duncan, Callum W.},
  journal = {Phys. Rev. B},
  volume = {109},
  issue = {1},
  pages = {014210},
  numpages = {11},
  year = {2024},
  month = {Jan},
  publisher = {American Physical Society},
  doi = {10.1103/PhysRevB.109.014210},
  url = {https://link.aps.org/doi/10.1103/PhysRevB.109.014210}
}

@article{Johnstone2021,
  title = {The mean-field Bose glass in quasicrystalline systems},
  volume = {54},
  ISSN = {1751-8121},
  url = {http://dx.doi.org/10.1088/1751-8121/ac1dc0},
  DOI = {10.1088/1751-8121/ac1dc0},
  number = {39},
  journal = {Journal of Physics A: Mathematical and Theoretical},
  publisher = {IOP Publishing},
  author = {Johnstone,  Dean and \"{O}hberg,  Patrik and Duncan,  Callum W},
  year = {2021},
  month = sep,
  pages = {395001}
}

@article{johnstoneMeanfieldPhasesUltracold2019,
  title = {Mean-Field Phases of an Ultracold Gas in a Quasicrystalline Potential},
  author = {Johnstone, Dean and {\"O}hberg, Patrik and Duncan, Callum W.},
  year = {2019},
  month = nov,
  journal = {Physical Review A},
  volume = {100},
  number = {5},
  pages = {053609},
  issn = {2469-9926, 2469-9934},
  doi = {10.1103/PhysRevA.100.053609},
  urldate = {2021-06-09},
  langid = {english}
}

@article{koschMultifrequencyOpticalLattice2022,
  title = {Multifrequency Optical Lattice for Dynamic Lattice-Geometry Control},
  author = {Kosch, M. N. and Asteria, L. and Zahn, H. P. and Sengstock, K. and Weitenberg, C.},
  year = {2022},
  month = nov,
  journal = {Physical Review Research},
  volume = {4},
  number = {4},
  pages = {043083},
  issn = {2643-1564},
  doi = {10.1103/PhysRevResearch.4.043083},
  urldate = {2023-09-06},
  langid = {english}
}

@article{krausFourDimensionalQuantumHall2013,
  title = {Four-{{Dimensional Quantum Hall Effect}} in a {{Two-Dimensional Quasicrystal}}},
  author = {Kraus, Yaacov E. and Ringel, Zohar and Zilberberg, Oded},
  year = {2013},
  month = nov,
  journal = {Physical Review Letters},
  volume = {111},
  number = {22},
  pages = {226401},
  doi = {10.1103/PhysRevLett.111.226401},
  urldate = {2020-05-01}
}

@article{maceQuantumSimulation2D2016,
  title = {Quantum {{Simulation}} of a {{2D Quasicrystal}} with {{Cold Atoms}}},
  author = {Mac{\'e}, Nicolas and Jagannathan, Anuradha and Duneau, Michel},
  year = {2016},
  month = sep,
  journal = {Crystals},
  volume = {6},
  number = {10},
  pages = {124},
  issn = {2073-4352},
  doi = {10.3390/cryst6100124},
  urldate = {2021-06-29},
  langid = {english}
}

@article{marraTopologicallyQuantizedCurrent2020,
  title = {Topologically Quantized Current in Quasiperiodic {{Thouless}} Pumps},
  author = {Marra, Pasquale and Nitta, Muneto},
  year = {2020},
  month = dec,
  journal = {Physical Review Research},
  volume = {2},
  number = {4},
  pages = {042035},
  issn = {2643-1564},
  doi = {10.1103/PhysRevResearch.2.042035},
  urldate = {2022-09-27},
  langid = {english}
}

@article{mirzhalilovPerpendicularSpaceAccounting2020,
  title = {Perpendicular Space Accounting of Localized States in a Quasicrystal},
  author = {Mirzhalilov, Murod and Oktel, M. {\"O}.},
  year = {2020},
  month = aug,
  journal = {Physical Review B},
  volume = {102},
  number = {6},
  pages = {064213},
  doi = {10.1103/PhysRevB.102.064213},
  urldate = {2022-01-19}
}

@article{oktelStrictlyLocalizedStates2021,
  title = {Strictly Localized States in the Octagonal {{Ammann-Beenker}} Quasicrystal},
  author = {Oktel, M. {\"O}.},
  year = {2021},
  month = jul,
  journal = {Physical Review B},
  volume = {104},
  number = {1},
  pages = {014204},
  doi = {10.1103/PhysRevB.104.014204},
  urldate = {2022-01-19}
}

@article{sbrosciaObservingLocalization2D2020,
  title = {Observing {{Localization}} in a {{2D Quasicrystalline Optical Lattice}}},
  author = {Sbroscia, Matteo and Viebahn, Konrad and Carter, Edward and Yu, Jr-Chiun and Gaunt, Alexander and Schneider, Ulrich},
  year = {2020},
  month = nov,
  journal = {Physical Review Letters},
  volume = {125},
  number = {20},
  pages = {200604},
  doi = {10.1103/PhysRevLett.125.200604},
  urldate = {2022-01-19}
}

@article{shechtmanMetallicPhaseLongRange1984,
  title = {Metallic {{Phase}} with {{Long-Range Orientational Order}} and {{No Translational Symmetry}}},
  author = {Shechtman, D. and Blech, I. and Gratias, D. and Cahn, J. W.},
  year = {1984},
  month = nov,
  journal = {Physical Review Letters},
  volume = {53},
  number = {20},
  pages = {1951--1953},
  issn = {0031-9007},
  doi = {10.1103/PhysRevLett.53.1951},
  urldate = {2019-11-24},
  langid = {english}
}

@article{gottlob2024quasiperiodicityprotectsquantizedtransport,
  title = {Quasiperiodicity Protects Quantized Transport in Disordered Systems Without Gaps},
  author = {Gottlob, Emmanuel and Borgnia, Dan S. and Slager, Robert-Jan and Schneider, Ulrich},
  journal = {PRX Quantum},
  volume = {6},
  issue = {2},
  pages = {020359},
  numpages = {13},
  year = {2025},
  month = {Jun},
  publisher = {American Physical Society},
  doi = {10.1103/zvng-w46m},
  url = {https://link.aps.org/doi/10.1103/zvng-w46m}
}

@article{Jagannathan2023,
  title = {Closing of gaps, gap labeling, and passage from molecular states to critical states in a two-dimensional quasicrystal},
  author = {Jagannathan, Anuradha},
  journal = {Phys. Rev. B},
  volume = {108},
  issue = {11},
  pages = {115109},
  numpages = {5},
  year = {2023},
  month = {Sep},
  publisher = {American Physical Society},
  doi = {10.1103/PhysRevB.108.115109},
  url = {https://link.aps.org/doi/10.1103/PhysRevB.108.115109}
}

@article{Griffiths1969,
  title = {Nonanalytic Behavior Above the Critical Point in a Random Ising Ferromagnet},
  volume = {23},
  ISSN = {0031-9007},
  url = {http://dx.doi.org/10.1103/PhysRevLett.23.17},
  DOI = {10.1103/physrevlett.23.17},
  number = {1},
  journal = {Physical Review Letters},
  publisher = {American Physical Society (APS)},
  author = {Griffiths,  Robert B.},
  year = {1969},
  month = jul,
  pages = {17–19}
}

@article{Agarwal2017,
  title = {Rare‐region effects and dynamics near the many‐body localization transition},
  volume = {529},
  ISSN = {1521-3889},
  url = {http://dx.doi.org/10.1002/andp.201600326},
  DOI = {10.1002/andp.201600326},
  number = {7},
  journal = {Annalen der Physik},
  publisher = {Wiley},
  author = {Agarwal,  Kartiek and Altman,  Ehud and Demler,  Eugene and Gopalakrishnan,  Sarang and Huse,  David A. and Knap,  Michael},
  year = {2017},
  month = jan 
}

@article{Johnstone2022,
  title = {Barriers to macroscopic superfluidity and insulation in a 2D Aubry–André model},
  volume = {55},
  ISSN = {1361-6455},
  url = {http://dx.doi.org/10.1088/1361-6455/ac6d34},
  DOI = {10.1088/1361-6455/ac6d34},
  number = {12},
  journal = {Journal of Physics B: Atomic,  Molecular and Optical Physics},
  publisher = {IOP Publishing},
  author = {Johnstone,  Dean and \"{O}hberg,  Patrik and Duncan,  Callum W},
  year = {2022},
  month = may,
  pages = {125302}
}

@article{strkaljCoexistenceLocalizationTransport2022,
  title = {Coexistence of Localization and Transport in Many-Body Two-Dimensional {{Aubry-Andr{\'e}}} Models},
  author = {{\v S}trkalj, Antonio and Doggen, Elmer V. H. and Castelnovo, Claudio},
  year = {2022},
  month = nov,
  journal = {Physical Review B},
  volume = {106},
  number = {18},
  pages = {184209},
  issn = {2469-9950, 2469-9969},
  doi = {10.1103/PhysRevB.106.184209},
  urldate = {2023-09-11},
  langid = {english}
}

@article{szaboMixedSpectraPartially2020,
  title = {Mixed Spectra and Partially Extended States in a Two-Dimensional Quasiperiodic Model},
  author = {Szab{\'o}, Attila and Schneider, Ulrich},
  year = {2020},
  month = jan,
  journal = {Physical Review B},
  volume = {101},
  number = {1},
  pages = {014205},
  issn = {2469-9950, 2469-9969},
  doi = {10.1103/PhysRevB.101.014205},
  urldate = {2020-03-12},
  langid = {english}
}

@article{szaboNonpowerlawUniversalityOnedimensional2018,
  title = {Non-Power-Law Universality in One-Dimensional Quasicrystals},
  author = {Szab{\'o}, Attila and Schneider, Ulrich},
  year = {2018},
  month = oct,
  journal = {Physical Review B},
  volume = {98},
  number = {13},
  pages = {134201},
  issn = {2469-9950, 2469-9969},
  doi = {10.1103/PhysRevB.98.134201},
  urldate = {2020-03-12},
  langid = {english}
}

@article{thielQuasicrystalsReachingMaturity1999a,
  title = {Quasicrystals. {{Reaching}} Maturity for Technological Applications},
  author = {Thiel, Patricia A. and Dubois, Jean-Marie},
  year = {1999},
  month = jan,
  journal = {Materials Today},
  volume = {2},
  number = {3},
  pages = {3--7},
  issn = {1369-7021},
  doi = {10.1016/S1369-7021(99)80058-3},
  urldate = {2024-04-13}
}

@article{uriSuperconductivityStrongInteractions2023,
  title = {Superconductivity and Strong Interactions in a Tunable Moir{\'e} Quasicrystal},
  author = {Uri, Aviram and {de la Barrera}, Sergio C. and Randeria, Mallika T. and {Rodan-Legrain}, Daniel and Devakul, Trithep and Crowley, Philip J. D. and Paul, Nisarga and Watanabe, Kenji and Taniguchi, Takashi and Lifshitz, Ron and Fu, Liang and Ashoori, Raymond C. and {Jarillo-Herrero}, Pablo},
  year = {2023},
  month = aug,
  journal = {Nature},
  volume = {620},
  number = {7975},
  pages = {762--767},
  issn = {1476-4687},
  doi = {10.1038/s41586-023-06294-z},
  urldate = {2024-04-13},
  copyright = {2023 The Author(s), under exclusive licence to Springer Nature Limited},
  langid = {english}
}

@article{verbinTopologicalPumpingPhotonic2015a,
  title = {Topological Pumping over a Photonic {{Fibonacci}} Quasicrystal},
  author = {Verbin, Mor and Zilberberg, Oded and Lahini, Yoav and Kraus, Yaacov E. and Silberberg, Yaron},
  year = {2015},
  month = feb,
  journal = {Physical Review B},
  volume = {91},
  number = {6},
  pages = {064201},
  doi = {10.1103/PhysRevB.91.064201},
  urldate = {2022-09-27}
}

@article{viebahnMatterWaveDiffractionQuasicrystalline2019,
  title = {Matter-{{Wave Diffraction}} from a {{Quasicrystalline Optical Lattice}}},
  author = {Viebahn, Konrad and Sbroscia, Matteo and Carter, Edward and Yu, Jr-Chiun and Schneider, Ulrich},
  year = {2019},
  month = mar,
  journal = {Physical Review Letters},
  volume = {122},
  number = {11},
  pages = {110404},
  issn = {0031-9007, 1079-7114},
  doi = {10.1103/PhysRevLett.122.110404},
  urldate = {2019-11-24},
  langid = {english}
}

@article{wallEffectiveManybodyParameters2015,
  title = {Effective Many-Body Parameters for Atoms in Nonseparable {{Gaussian}} Optical Potentials},
  author = {Wall, Michael L. and Hazzard, Kaden R. A. and Rey, Ana Maria},
  year = {2015},
  month = jul,
  journal = {Physical Review A: Atomic, Molecular, and Optical Physics},
  volume = {92},
  number = {1},
  pages = {013610},
  doi = {10.1103/PhysRevA.92.013610}
}

@article{weiHubbardParametersProgrammable2024,
  title = {Hubbard Parameters for Programmable Tweezer Arrays},
  author = {Wei, Hao-Tian and {Ibarra-Garc{\'{\i}}a-Padilla}, Eduardo and Wall, Michael L. and Hazzard, Kaden R. A.},
  year = {2024},
  month = jan,
  journal = {Physical Review A: Atomic, Molecular, and Optical Physics},
  volume = {109},
  number = {1},
  pages = {013318},
  doi = {10.1103/PhysRevA.109.013318}
}

@article{yaoEdgeStatesTopological2018,
  title = {Edge States and Topological Invariants of Non-Hermitian Systems},
  author = {Yao, Shunyu and Wang, Zhong},
  year = {2018},
  month = aug,
  journal = {Physical Review Letters},
  volume = {121},
  number = {8},
  eprint = {https://www.pnas.org/doi/pdf/10.1073/pnas.1720865115},
  pages = {086803},
  doi = {10.1103/PhysRevLett.121.086803}
}

@article{zhuConstructionMaximallyLocalized2017,
  title = {Construction of {{Maximally Localized Wannier Functions}}},
  author = {Zhu, Junbo and Chen, Zhu and Wu, Biao},
  year = {2017},
  month = oct,
  journal = {Frontiers of Physics},
  volume = {12},
  number = {5},
  eprint = {1609.05992},
  pages = {127102},
  issn = {2095-0462, 2095-0470},
  doi = {10.1007/s11467-016-0628-8},
  urldate = {2021-12-17},
  archiveprefix = {arXiv},
  langid = {english}
}

@article{zhuThermodynamicPhaseDiagram2023,
  title = {Thermodynamic Phase Diagram of Two-Dimensional Bosons in a Quasicrystal Potential},
  author = {Zhu, Zhaoxuan and Yao, Hepeng and {Sanchez-Palencia}, Laurent},
  year = {2023},
  month = may,
  journal = {Physical Review Letters},
  volume = {130},
  number = {22},
  pages = {220402},
  doi = {10.1103/PhysRevLett.130.220402}
}

@misc{jagannathan2024propertiesammannbeenkertilingsquare,
      title={Properties of the Ammann-Beenker tiling and its square approximants}, 
      author={Anuradha Jagannathan and Michel Duneau},
      year={2024},
      eprint={2308.07701},
      archivePrefix={arXiv},
      primaryClass={cond-mat.str-el},
      url={https://arxiv.org/abs/2308.07701}, 
}

@article{ShklovskiiStat1993,
  title = {Statistics of spectra of disordered systems near the metal-insulator transition},
  author = {Shklovskii, B. I. and Shapiro, B. and Sears, B. R. and Lambrianides, P. and Shore, H. B.},
  journal = {Phys. Rev. B},
  volume = {47},
  issue = {17},
  pages = {11487--11490},
  numpages = {0},
  year = {1993},
  month = {May},
  publisher = {American Physical Society},
  doi = {10.1103/PhysRevB.47.11487},
  url = {https://link.aps.org/doi/10.1103/PhysRevB.47.11487}
}

@article{HanCritical1994,
  title = {Critical and bicritical properties of Harper's equation with next-nearest-neighbor coupling},
  author = {Han, J. H. and Thouless, D. J. and Hiramoto, H. and Kohmoto, M.},
  journal = {Phys. Rev. B},
  volume = {50},
  issue = {16},
  pages = {11365--11380},
  numpages = {0},
  year = {1994},
  month = {Oct},
  publisher = {American Physical Society},
  doi = {10.1103/PhysRevB.50.11365},
  url = {https://link.aps.org/doi/10.1103/PhysRevB.50.11365}
}

@misc{8QCHamiltonian,
author = {Emmanuel Gottlob},
title = {Large-Scale Tight-Binding Hamiltonians of the Eightfold Quasicrystalline Potential (8QC)},
doi = {10.17863/CAM.111974},
howpublished= {\url{https://doi.org/10.17863/CAM.95664}}}

@article{YouFibonacci1991,
doi = {10.1088/0953-8984/3/38/003},
url = {https://dx.doi.org/10.1088/0953-8984/3/38/003},
year = {1991},
month = {sep},
publisher = {},
volume = {3},
number = {38},
pages = {7255},
author = {J Q You and  J R Yan and  Tiansheng Xie and  Xiaobiao Zeng and  J X Zhong},
title = {Generalized Fibonacci lattices: dynamical maps, energy spectra and wavefunctions},
journal = {Journal of Physics: Condensed Matter}
}

@article{OstlundSchrodinger1983,
  title = {One-Dimensional Schr\"odinger Equation with an Almost Periodic Potential},
  author = {Ostlund, Stellan and Pandit, Rahul and Rand, David and Schellnhuber, Hans Joachim and Siggia, Eric D.},
  journal = {Phys. Rev. Lett.},
  volume = {50},
  issue = {23},
  pages = {1873--1876},
  numpages = {0},
  year = {1983},
  month = {Jun},
  publisher = {American Physical Society},
  doi = {10.1103/PhysRevLett.50.1873},
  url = {https://link.aps.org/doi/10.1103/PhysRevLett.50.1873}
}

@article{DevakulAnderson2017,
  title = {Anderson localization transitions with and without random potentials},
  author = {Devakul, Trithep and Huse, David A.},
  journal = {Phys. Rev. B},
  volume = {96},
  issue = {21},
  pages = {214201},
  numpages = {9},
  year = {2017},
  month = {Dec},
  publisher = {American Physical Society},
  doi = {10.1103/PhysRevB.96.214201},
  url = {https://link.aps.org/doi/10.1103/PhysRevB.96.214201}
}

@article{ZhuLocalization2024,
  title = {Localization and spectral structure in two-dimensional quasicrystal potentials},
  author = {Zhu, Zhaoxuan and Yu, Shengjie and Johnstone, Dean and Sanchez-Palencia, Laurent},
  journal = {Phys. Rev. A},
  volume = {109},
  issue = {1},
  pages = {013314},
  numpages = {21},
  year = {2024},
  month = {Jan},
  publisher = {American Physical Society},
  doi = {10.1103/PhysRevA.109.013314},
  url = {https://link.aps.org/doi/10.1103/PhysRevA.109.013314}
}
\appendix

\subsection{Sinc Discrete Variable representation} \label{app:sincdvr}

The slow, polynomial, convergence of finite-difference (FD) methods calls for the use of alternatives with improved convergence rate. This can be achieved with the use of discrete variable representation (DVR) methods, which have been commonly used in chemical and molecular physics, but did not find their way to the cold atoms community until recently \citep{wallEffectiveManybodyParameters2015, weiHubbardParametersProgrammable2024}. We review here the sinc DVR method, which we used to replace the finite-difference method previously implemented in our code for constructing Wannier functions, and which allowed to alleviate the memory usage by a factor of 25 -- permitting  to vastly increase the maximum system size compared to what is achievable with the FD method. 

While the finite-difference method introduced earlier uses a basis of rectangular functions centered around the grid points, the sinc DVR methods relies on a basis of sinc functions:
\begin{equation}
    \braket{x}{\Delta_n} = \frac{1}{\sqrt{\delta x}} \text{sinc}\left( \pi (x-x_n)/\delta x\right) \,.
\end{equation}

This discrete basis set can be interpreted as a ensemble of low-passed delta functions, with a momentum cut-off $K = \pi/\delta x$, and provides the significant advantage of an exponential convergence rate with the inverse grid spacing for functions that do not contain momenta beyond $K$ \citep{wallEffectiveManybodyParameters2015}.

The sinc DVR basis set is orthonormal, as can be computed with the integral:
\begin{align}
    \int dx \braket{\Delta_n}{x} \braket{x}{\Delta_m} = \delta_{n,m} \,.
\end{align}

The kinetic energy operator expressed in the sinc DVR basis can be similarly obtained as:
\begin{align}
T_{nn'} &\equiv \left\langle \Delta_n \middle| -\frac{\hbar^2}{2m} \frac{\partial^2}{\partial x^2} \middle| \Delta_{n'} \right\rangle \nonumber \\
&= \frac{\hbar^2}{2m \delta x^2} \begin{cases}
\frac{\pi^2}{3}, & n = n' \\
2(-1)^{n-n'}\frac{1}{(n-n')^2}, & \text{otherwise}.
\end{cases}
\end{align}

and the potential energy operator is exactly diagonal for potentials $V(x)$ that do not contain frequencies higher than the cut-off frequency, or approximately diagonal (with an error exponentially small in $K$) in the case where $V(x)$ contains frequencies beyond $K$:
\begin{equation}
    V_{nn'}\approx V(x_n) \delta_{nn'}\,.
\end{equation}

Compared to the finite-difference method, the kinetic operator loses sparsity, but still remains block diagonal along each direction in dimensions higher than $1$. This decreased sparsity increases the time required to diagonalize $H_{cont}$, however this is generally entirely counterbalanced by the exponential convergence, which allows to use orders of magnitude less grid points. In our case, we found that the DVR method could reach the level of convergence obtained by finite-difference methods (for onsite energies and tunneling amplitudes) with $5$ times less grid points per dimension, resulting in a 25-fold decrease in the number of grid points over both dimensions. This allows to vastly reduce the memory required for storing the Wannier functions.

We used the methods described in \citep{gottlobHubbardModelsQuasicrystalline2023}, updated with the sinc DVR method, to generate the large scale tight-binding Hamiltonian, on a circular patch of the 8QC of diameter $70 \lambda$ for lattice depths between $1.0$ and $6.0\ E_r$, with $12479$ lattice sites. The lattice depths were generated in steps of $0.125 \ E_r$, and a finer spacing of lattice depths was then achieved by linearly interpolating the onsite energies and tunneling elements between the numerically computed steps. The parameters used for this dataset were a real-space grid spacing $\delta_x = 0.1 \lambda$, and a cutoff radius for generating the Wannier functions of $R = 4 \lambda$. To reflect experimental conditions, we set phases $\phi_i$ to randomly selected values:
\begin{equation}
    \begin{pmatrix}
        \phi_1\\ \phi_2 \\ \phi_3 \\ \phi_4
    \end{pmatrix}
=
    \begin{pmatrix}
        -469.66810\\ -772.74425\\  -879.16884\\  215.87799
    \end{pmatrix}
     \text{mod} \  2 \pi \,, \nonumber
\end{equation}
 which translates the center of the system to the arbitrary chosen location $$(\Delta_x, \Delta_y)= \left(74.75 \ \lambda, 122.98607
 \ \lambda\right).$$ This ensures that the center of global eight-fold symmetry is located outside the considered system. The obtained tight-binding Hamiltonians are publicly available on \citep{8QCHamiltonian}.

\end{document}